\documentclass[twocolumn]{aastex631}
\usepackage{comment}
\usepackage{amsmath}	
\usepackage{amssymb}
\usepackage{footnote}
\usepackage{appendix}
\usepackage{upgreek}
\usepackage{color, colortbl}
\usepackage{tablefootnote}
\usepackage[graphicx]{realboxes}
\usepackage{threeparttable}
\usepackage[caption=false]{subfig}
\usepackage{gensymb}
\usepackage{xcolor,colortbl}

\definecolor{Gray}{gray}{0.9}
\graphicspath{{./}{figures/}}

\newcounter{daggerfootnote}

\newcommand{\rf}{{\sc Realfast}}

\newcommand{\mhalflight}{\phi} 
\newcommand{\halflight}{$\mhalflight$}

\newcommand{\mgmag}{m}
\newcommand{\gmag}{$\mgmag$}

\newcommand{\mpchance}{P^c}
\newcommand{\pchance}{$\mpchance$}

\newcommand{\mPO}{P(O)}  
\newcommand{\PO}{$\mPO$}

\newcommand{\mPOx}{P(O|x)}  
\newcommand{\POx}{$\mPOx$}

\shorttitle{FRB host galaxies}
\shortauthors{Bhandari et al.}
\received{November 10, 2021}
\submitjournal{ApJ}

\begin{document}
\title{Characterizing the FRB host galaxy population and its connection to transients in the local and extragalactic Universe}

\correspondingauthor{Shivani Bhandari}
\email{shivani.bhandari@csiro.au}
\author[0000-0003-3460-506X]{Shivani Bhandari}
\affil{CSIRO, Space and Astronomy, PO Box 76, Epping NSW 1710 Australia}
\author{Kasper E. Heintz}
\affil{Centre for Astrophysics and Cosmology, Science Institute, University of Iceland, Dunhagi 5, 107 Reykjav\'ik, Iceland}
\affil{Cosmic Dawn Center (DAWN), Niels Bohr Institute, University of Copenhagen, Jagtvej 128, DK-2100 Copenhagen \O, Denmark}
\author[0000-0002-2059-0525]{Kshitij Aggarwal}
\affil{Department of Physics and Astronomy, West Virginia University, P.O. Box 6315, Morgantown, WV 26506, USA}
\affil{Center for Gravitational Waves and Cosmology, West Virginia University, Chestnut Ridge Research Building, Morgantown, WV 26505, USA}
\author[0000-0003-1483-0147]{Lachlan Marnoch}
\affiliation{Department of Physics and Astronomy, Macquarie University, NSW 2109, Australia}
\affil{CSIRO, Space and Astronomy, PO Box 76, Epping NSW 1710 Australia}
\affil{Astronomy, Astrophysics and Astrophotonics Research Centre, Macquarie University, Sydney, NSW 2109, Australia}
\author{Cherie K. Day}
\affil{Centre for Astrophysics and Supercomputing, Swinburne University of Technology, John St, Hawthorn, VIC 3122, Australia}
\affil{CSIRO, Space and Astronomy, PO Box 76, Epping NSW 1710 Australia}
\author{Jessica Sydnor}
\affil{Department of Physics and Astronomy, West Virginia University, P.O. Box 6315, Morgantown, WV 26506, USA}
\affil{Center for Gravitational Waves and Cosmology, West Virginia University, Chestnut Ridge Research Building, Morgantown, WV 26505, USA}
\author{Sarah Burke-Spolaor}
\affil{Department of Physics and Astronomy, West Virginia University, P.O. Box 6315, Morgantown, WV 26506, USA}
\affil{Center for Gravitational Waves and Cosmology, West Virginia University, Chestnut Ridge Research Building, Morgantown, WV 26505, USA}
\affil{Canadian Institute for Advanced Research, CIFAR Azrieli Global Scholar, MaRS Centre West Tower, 661 University Ave. Suite 505, Toronto ON M5G 1M1, Canada}
\author[0000-0002-4119-9963]{Casey J.~Law}
\affil{Cahill Center for Astronomy and Astrophysics, MC 249-17 California Institute of Technology, Pasadena, CA 91125, USA}
\affiliation{Owens Valley Radio Observatory, California Institute of Technology, 100 Leighton Lane, Big Pine, CA, 93513, USA}
\author[0000-0002-7738-6875]{J. Xavier Prochaska}
\affil{University of California, Santa Cruz, 1156 High St., Santa Cruz, CA 95064, USA}
\affiliation{
Kavli Institute for the Physics and Mathematics of the Universe (Kavli IPMU),
5-1-5 Kashiwanoha, Kashiwa, 277-8583, Japan}
\author{Nicolas Tejos}
\affil{Instituto de F\'isica, Pontificia Universidad Cat\'olica de Valpara\'iso, Casilla 4059, Valpara\'iso, Chile}

\author[0000-0003-2149-0363]{Keith W. Bannister}
\affil{CSIRO, Space and Astronomy, PO Box 76, Epping NSW 1710 Australia}
\author{Bryan J. Butler}
\affil{National Radio Astronomy Observatory, Socorro, NM, USA}
\author{Adam T.Deller}
\affiliation{Centre for Astrophysics and Supercomputing, Swinburne University of Technology, John St, Hawthorn, VIC 3122, Australia}
\author{R.~D.~Ekers}
\affil{CSIRO, Space and Astronomy, PO Box 76, Epping NSW 1710 Australia}

\author{Chris Flynn}
\affiliation{Centre for Astrophysics and Supercomputing, Swinburne University of Technology, John St, Hawthorn, VIC 3122, Australia}
\author[0000-0002-7374-935X]{Wen-fai Fong}
\affil{Center for Interdisciplinary Exploration and Research in Astrophysics (CIERA) and Department of Physics and Astronomy, Northwestern University, Evanston, IL 60208, USA}
\author{Clancy~W.~James}
\affiliation{International Centre for Radio Astronomy Research, Curtin University, Bentley, WA 6102, Australia}
\author{T.~Joseph~W.~Lazio}
\affil{Jet Propulsion Laboratory, California Institute of Technology}
\author[0000-0002-4300-121X]{Rui Luo}
\affil{CSIRO, Space and Astronomy, PO Box 76, Epping NSW 1710 Australia}
\author{Elizabeth K. Mahony}
\affil{CSIRO, Space and Astronomy, PO Box 76, Epping NSW 1710 Australia}
\author[0000-0003-4501-8100]{Stuart D. Ryder}
\affil{Department of Physics and Astronomy, Macquarie University, NSW 2109, Australia}
\affil{Astronomy, Astrophysics and Astrophotonics Research Centre, Macquarie University, Sydney, NSW 2109, Australia}
\author{Elaine M. Sadler}
\affil{CSIRO, Space and Astronomy, PO Box 76, Epping NSW 1710 Australia}
\affil{Sydney Institute for Astronomy, School of Physics A28, The University of Sydney, NSW 2006, Australia}
\author[0000-0002-7285-6348]{Ryan M. Shannon}
\affiliation{Centre for Astrophysics and Supercomputing, Swinburne University of Technology, John St, Hawthorn, VIC 3122, Australia}
\author[0000-0002-9274-3092]{JinLin Han}
\affiliation{National Astronomical Observatories, Chinese Academy of Sciences, 20A Datun Road, Chaoyang District, Beijing 100012, China}
\affiliation{CAS Key Laboratory of FAST, NAOC, Chinese Academy of Sciences, Beijing 100101, China}
\affiliation{School of Astronomy, University of Chinese Academy of Sciences, Beijing 100049, China}
\author{Kejia Lee}
\affiliation{Kavli Institute for Astronomy and Astrophysics, Peking University, Beijing 100871, P.R.China}
\affiliation{National Astronomical Observatories, Chinese Academy of Sciences, 20A Datun Road, Chaoyang District, Beijing 100012, China}
\author[0000-0002-9725-2524]{Bing Zhang}
\affiliation{Department of Physics and Astronomy, University of Nevada, Las Vegas, Las Vegas, NV 89154, USA}

\begin{abstract}
We present the localization and host galaxies of one repeating and two apparently non-repeating Fast Radio Bursts. FRB\,20180301A was detected and localized with the Karl G. Jansky Very Large Array to a star-forming galaxy at $z=0.3304$. FRB20191228A, and FRB20200906A were detected and localized by the Australian Square Kilometre Array Pathfinder to host galaxies at $z=0.2430$ and $z=0.3688$, respectively. We combine these with 13 other well-localized FRBs in the literature, and analyze the host galaxy properties. We find no significant differences in the host properties of repeating and apparently non-repeating FRBs. FRB hosts are moderately star-forming, with masses slightly offset from the star-forming main-sequence. Star formation and low-ionization nuclear emission-line region (LINER) emission are major sources of ionization in FRB host galaxies, with the former dominant in repeating FRB hosts. FRB hosts do not track stellar mass and star formation as seen in field galaxies (more than 95\% confidence). FRBs are rare in massive red galaxies, suggesting that progenitor formation channels are not solely dominated by delayed channels which lag star formation by Gigayears. The global properties of FRB hosts are indistinguishable from core-collapse supernovae (CCSNe) and short gamma-ray bursts (SGRBs) hosts, and the spatial offset (from galaxy centers) of FRBs is mostly inconsistent with that of the Galactic neutron star population (95\% confidence). The spatial offsets of FRBs (normalized to the galaxy effective radius) also differ from those of globular clusters (GCs) in late- and early-type galaxies with 95\% confidence.

\end{abstract}


\section{Introduction} \label{intro}
The physical mechanism and the source population(s) powering the energetic ($10^{-2}$ to $\sim 400$~Jy\,ms) and $\upmu$s-ms duration fast radio bursts (FRBs) are currently two of the biggest mysteries of modern astrophysics. Significant progress has been made in the last two years, with (sub-)arcsecond localization of FRBs using radio interferometers such as the Karl G. Jansky Very Large Array (VLA), the Australian Square Kilometre Array Pathfinder (ASKAP), the Deep Synoptic Array (DSA-10), and the European VLBI Network (EVN) \citep{VLAlocalisation, Bannister+19, Ravi+19a, Marcote2020,Law2020}. These FRB localizations have enabled the identification of FRB host galaxies, and in some cases the location of the FRBs within the galaxies, which has provided the clues to the progenitor channels of these enigmatic sources. While the nature of FRBs remains uncertain, the redshifts for localized FRBs still permit us to probe the baryonic content of the Universe \citep{JP+20}, through measurements of the ionised baryon density in the intergalactic medium (IGM). FRBs are thus excellent probes of cosmology and the structure of the Universe \citep{Bhandari+21}.

Recently, the Canadian Hydrogen Intensity Mapping Experiment Fast Radio Burst (CHIME/FRB) Project published a catalog of 535 FRBs detected during their first year of operations and observed in the range $400-800$~MHz \citep{CHIMECatalog}. The catalog represents the first  large sample ($\gg 10^2$) of FRBs detected in a well-controlled experiment. It includes bursts from repeating and non-repeating FRB sources, and facilitates a comparative study of the FRB population. The CHIME/FRB sample reveals that while the sky locations and dispersion measures (DMs) of the two FRB populations are consistent with being drawn from the same distribution, the bursts from repeating sources have wider pulse widths and narrower bandwidths compared to apparent non-repeaters \citep{Fonseca+20,Pleunis+21}. Thus, these observational differences in the bursts themselves hint at different propagation or emission mechanisms for the two FRB sub-populations. Whether all FRBs repeat is still an outstanding question in the field \citep{Palaniswamy+18,James+19,Caleb+19,Ravi+19,Shunke+21}.

In addition to the properties of the bursts, analyses of their host galaxy environments may provide more clues to disentangle the two FRB populations and their progenitor sources \citep{Chittidi2020, Tendulkar+21}. Currently, 15 FRBs (including repeating and non-repeating) have been localized to host galaxies at redshifts in the range $z = 0.03-0.66$ \citep{Heintz+20}\footnote{\url{http://frbhosts.org}}. This preliminary sample shows that FRB host galaxies overall exhibit a broad range of color ($u-r = 0.9 - 2.0$), stellar mass (M$_\star = 10^{8} - 10^{10}\,$M$_{\odot}$), and star-formation rates (${\rm SFR} = 0.05 - 10\,$M$_{\odot}\,{\rm yr}^{-1}$). Moreover, the burst sites are in most cases significantly offset from the host-galaxy centers. A high spatial resolution analysis of a subset of FRB hosts found that most FRBs are not located in regions of elevated local star formation and stellar mass surface densities in comparison to the mean global values of their hosts \citep{Mannings+20}. Also, the majority of hosts in \citet{Mannings+20} show clear spiral arm features in the infrared (IR), with the positions of the bursts found to be consistent with an origin in the spiral arms. An analysis of the host-burst offset distribution and other host properties rule out long gamma-ray bursts (LGRBs) and super-luminous supernovae (SLSNe) as FRB production channels and favor compact merger events (double white dwarf (WD) and double neutron star (NS) mergers), accretion-induced collapse (AIC) of a white dwarf and core-collapse supernovae (CCSNe) to be plausible mechanisms for non-repeating ASKAP-localized bursts \citep{Bhandari_2020a,Heintz+20,Ye+20}.

In this first, relatively small host galaxy sample, tentative evidence was found for the hosts of the repeating FRBs to be less massive and less luminous on average, compared to the hosts of the apparently non-repeating FRBs. Even within the specific sub-population of repeating FRB hosts, though, a very diverse nature is apparent; 
FRB20121102A originates in a highly magnetized environment \citep{Michilli2018} co-located with a radio nebula in a low-metallicity, highly star-forming dwarf galaxy at $z=0.1927$ \citep{VLAlocalisation,Tendulkar2016}. The immediate environment and host galaxy properties of the FRB20121102A source led to a concordant model for FRBs in which bursts are produced by young magnetars, remnants from SLSNe or LGRBs \citep{Margalit+2018}. In contrast, FRB20180916B originates close to a star-forming region in a massive nearby ($z=0.03$) spiral galaxy \citep{Marcote2020} lacking an extreme magneto-ionic environment and radio nebula. Observations with the \textit{Hubble Space Telescope} (HST) established a small but significant offset of
FRB20180916B from the nearest knot of active star formation in the host galaxy. This suggests that the age of the progenitor is inconsistent with that of a young magnetar but compatible with the ages of high-mass X-ray binaries and gamma-ray binaries \citep{Tendulkar+21}. Another repeating source, FRB20200120E is localized to a globular cluster system in the nearby spiral galaxy M81, suggesting that if the progenitor is a young neutron star, it must have been formed via an alternative pathway, such as accretion-induced collapse of a white dwarf, or the merger of compact stars in a binary system \citep{Bhardwaj+21,Kirsten+21}. Finally, the repeating FRBs\,20190711A and 20201124A are observed to originate in typical star-forming galaxies at their respective redshifts \citep{Heintz+20,Ravi+21,Fong+21}. 

In this paper, we introduce three additional bursts and their host galaxies to the existing sample and conduct a differential study of the host population of one-off and repeating FRBs using an updated sample of 16 FRB hosts (containing 6 repeating and 10 non-repeating FRBs). The paper is laid out as follows: Section \ref{sec:reploc} presents the discovery of a new burst from the repeating FRB20180301A source \citep{Luo+20Nat} and resulting localization using the {\rf} system at the VLA. We also describe the radio and optical properties of the host galaxy of FRB20180301A. Section \ref{sec:CRAFT_FRBs} presents the discovery and localization of two new (apparently) non-repeating FRBs, namely FRB20191228A and FRB20200906A, found in the Commensal Real-time ASKAP Fast Transients (CRAFT) Survey \citep{mbb+10}. We also describe the follow-up observations and identification of their host galaxies. In Section \ref{sec:FRBs_galaxies}, we compare the overall population of FRB hosts with field galaxies at similar redshifts, based on the largest sample to date. In Section \ref{sec:rep_oneoff}, we differentiate the hosts and compare the properties of the two sub-populations; repeating and one-off bursts. In Section \ref{sec:galactic}, we compare the global properties and the projected physical offsets of FRBs to a range of both extragalactic transients and the projected source population of Galactic objects. Additionally, we compare the host-normalized offset distributions of FRBs with that of globular clusters associated with different galaxy types. We summarise our results in Section \ref{sec:summary}. 

\begin{figure}
\includegraphics[scale=0.39]{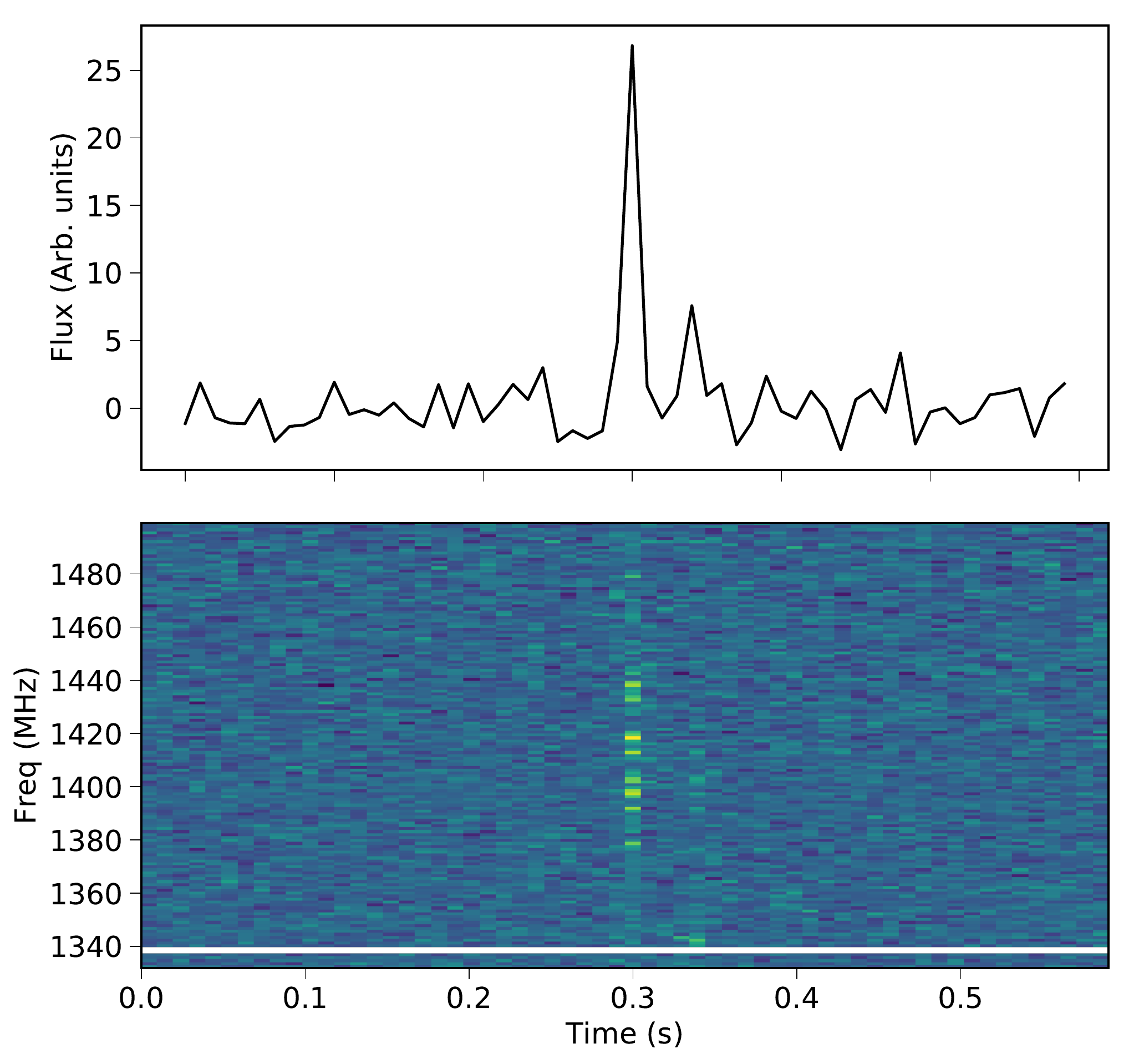} 
\includegraphics[scale=0.42,angle=0,clip,trim={3.5cm 0cm 0.0cm 0.5cm}]{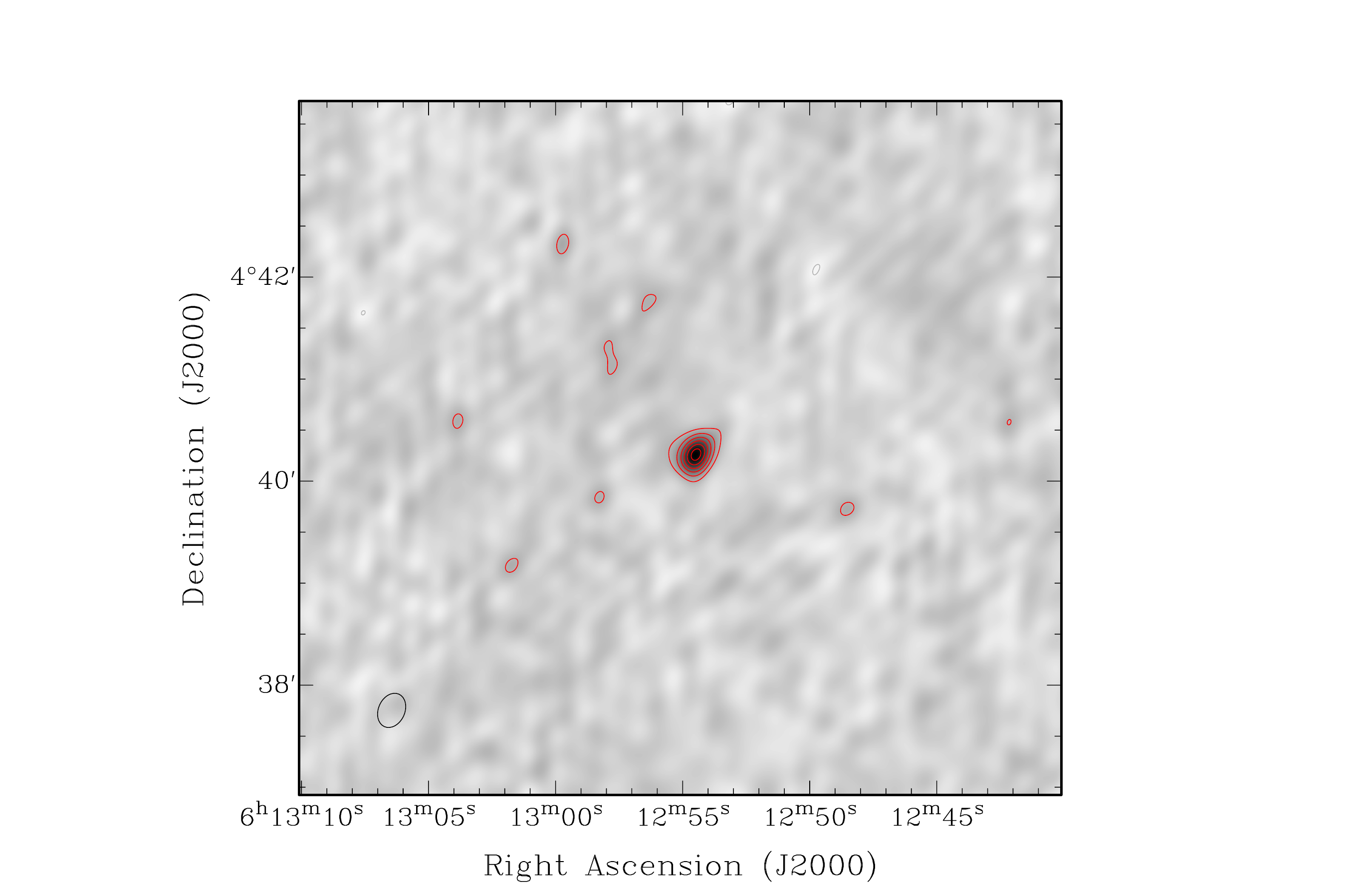}
\caption{\rf\ detection of FRB20180301A. Top and middle panels show the dispersion-corrected frequency integrated profile and spectrum of FRB20180301A. Bottom panel shows the fast sampled (10\,ms) calibrated radio image of FRB20180301A. The red contour levels are placed at 11.2\,mJy\,beam$^{-1}$ times factors of [$-3$, 3, 6, 9, 12, 15, 18]. The synthesised beam in the burst image is $20.9\arcsec \times 15.5\arcsec$ at a position angle of $-24.1\degree$.}
\label{fig:realfast}
\end{figure}
\section{Localization of the repeating FRB20180301A} \label{sec:reploc}

\subsection{Detection of a repeating burst}
FRB20180301A, a burst originally detected in the Breakthrough Listen project at the Parkes radio telescope \citep{Price2019}, was observed to emit repeating radio pulses in sensitive follow-up observations with the FAST telescope using the 19-beam receiver centered at 1.25\,GHz \citep{Luo+20Nat}. Fifteen repeating bursts with fluences ranging from 0.03~Jy\,ms to 0.4~Jy\,ms, were detected from the source of FRB20180301A in a total effective observation time of $\sim$12~hrs at FAST. 

We performed follow-up observations of FRB20180301A using Director's Discretionary Time (DDT) with 40\,hrs scheduled under VLA/19B-351, using the VLA in C array configuration at L-band, spanning 1-2\,GHz. We used the L16f5DC-realfast correlator mode, which enables a fast sampling time of 10\,ms. We observed the field of FRB20180301A centered at RA(J2000): 06:12:54.96 and Dec(J2000): +04:38:43.60 covering a localization uncertainty region of 2.6$'$ at 1.25\,GHz obtained from detection at FAST. The source was observed at 9 epochs, each with an observation time of 2\,hr in the period Feb$-$May 2020. 
Each observation had an on-source time of $\sim$1.5\,hrs and was searched for fast transients by the \rf\ system. The details of the \rf\ search procedure can be found in \citet{Law2018,Law2020}.
We detected a repeat burst from FRB20180301A on 2020 May 28 at UTC 19:14:47.310 in the last epoch. 
It was detected in the real-time system with an image S/N of 10.8 at a DM of 517.45\,pc\,cm$^{-3}$. The Deep Learning based classifier \texttt{fetch} \citep{fetch} reported an astrophysical probability of 99\% for this burst. We used the recorded visibilities of the burst data to refine our estimate of the burst properties in an offline reprocessing. We re-ran the search with a finer DM grid at $0.1\%$ fractional sensitivity loss on a sub-band of the data. This sub-band was manually identified to consist of burst signal. The refined search led to an improvement in detection significance to $\sim 19$ at the DM of 536\,pc\,cm$^{-3}$. The top panel of Fig.\,\ref{fig:realfast} shows the burst profile and dedispersed spectrogram. This figure also shows a weaker component $\sim50$\,ms after the burst that is present in the lower part of the band. This could either be a component of the main burst, or another burst from this FRB. 
We used the visibilities from the main component to determine the localization of this FRB. 

We used \textsc{burstfit}\footnote{\url{https://github.com/thepetabyteproject/burstfit}} \citep{2021arXiv210705658A} to model the spectro-temporal properties of FRB20180301A using its spectrogram. Following the method described in \citet{2021arXiv210705658A}, we modelled the spectrum and the pulse using a Gaussian, and fit for the following parameters simultaneously: arrival time of peak, Gaussian FWHM of pulse, peak of the spectrum, FWHM of the spectrum, fluence and DM. We used \textsc{scipy.curve\_fit} followed by Markov-Chain Monte-Carlo methods to obtain the posterior distribution of all the fit parameters. The resulting burst properties are presented in Table\,\ref{tab:measurements}.

\begin{deluxetable*}{lccc}
\tablewidth{0pc}
\tablecaption{Measured and derived properties of the new FRBs presented in this work. \label{tab:measurements}}
\tabletypesize{\footnotesize}
\tablehead{\colhead{Properties}\tablenotemark{} & \colhead{FRB20180301A} &\colhead{FRB20191228A}  
& \colhead{FRB20200906A} 
\\
} 
\startdata 
         Arrival time\tablenotemark{a} (UT)  & 2020-05-28 19:14:47.310 & 2019-12-28 09:16:16.444 & 2020-09-06 21:40:50.923 \\
         S/N  & 19.0 \tablenotemark{b} & 22.9 & 19.2 \\
         DM (pc\,cm$^{-3}$) & $536^{+8}_{-13}$ & $297.5\pm 0.05$ & $577.8\pm 0.02$\\
         DM$_{\rm ISM}~\rm NE2001 $ (pc\,cm$^{-3}$) & 152 & 33 & 36\\
         DM$_{\rm ISM}~\rm YMW16 $ (pc\,cm$^{-3}$) & 254 & 20 & 38\\
         DM$_{\rm cosmic}$\tablenotemark{c} (pc\,cm$^{-3}$) & 289  & 210 & 324 \\
         RA (J2000) & 06h12m54.44s\,($\pm0.01\arcsec$$\pm0.62\arcsec$) & 22h57m43.30s\,($\pm0.34\arcsec$$\pm0.83\arcsec$)  & 03h33m59.08s\,($\pm0.10\arcsec$$\pm0.34\arcsec$) \\ 
         Dec (J2000) & +04d40$'$15.8\,($\pm0.01\arcsec$$\pm0.6\arcsec$) &  $-$29d35$'$38.7\,($\pm0.3\arcsec$$\pm0.8\arcsec$) & $-$14d$04'59.5$\,($\pm0.1\arcsec$$\pm0.6\arcsec$) \\
         Fluence (Jy\,ms) & $4.9^{+0.5}_{-0.4}$ & 40$^{+100}_{-40}$ & 59$^{+25}_{-10}$ \\
         Pulse width (ms)& $7^{+2}_{-3}$ & $2.3\pm 0.6$ \tablenotemark{d} & $6.0\pm 0.6$ \\
         Spectral energy density\tablenotemark{e} (erg\,Hz$^{-1}$)   & $6.9 \times 10^{30}$ & $6.4 \times 10^{31}$ & $1.1 \times 10^{32}$ \\
         Persistent source, & $<1.8\times 10^{22}$ & $<3.4\times 10^{21}$ & $<4.3\times 10^{21}$  \\
         radio luminosity (W\,Hz$^{-1}$) & 1.5\,GHz & 6.5\,GHz & 6\,GHz\\
         \\
         \textbf{Host galaxy} \\
        Redshift\tablenotemark{*} & 0.3304 & 0.2432 & 0.3688 \\
        $u-r$\tablenotemark{$\dagger$} (restframe) & $0.90 \pm0.11$ & $2.13 \pm0.74$ & $1.22\pm 0.11$ \\
        M$_{\rm r}$\tablenotemark{$\dagger$} (restframe) & $-20.18\pm 0.07$ & $-18.26\pm0.05$ &  $-21.49 \pm 0.05$ \\
        Galactic $E(B-V)$\tablenotemark{$\dagger$} & 0.46 & 0.02 & 0.05 \\
        M$_*$\tablenotemark{$\dagger$} ($10^{10}~M_\odot$) & $0.23\pm 0.06$ &$0.54\pm 0.60$ & $1.33 \pm 0.37$ \\
        SFR\tablenotemark{*} ($M_\odot$~yr$^{-1}$) & $1.93\pm0.58$ & $0.50\pm 0.15$& $0.48 \pm 0.14$\\
        log(sSFR) (yr$^{-1}$) & $-$9.08 & $-$10.03 & $-$10.44\\
        Metallicity\tablenotemark{f} \tablenotemark{*} & 8.70 & 8.48 & 8.76 \\
        Projected offset from galaxy center (kpc) & $10.8 \pm3.0$ & $5.7\pm3.3$ & $5.9\pm2.0$ \\
        Effective radius (kpc) & $5.80\pm 0.20$ &$1.78\pm0.06$  & $7.58 \pm 0.06$ \\
 \tableline
\enddata 
\tablenotetext{a}{Arrival time of a repeating burst from FRB20180301A source and one-off ASKAP/CRAFT FRBs.}
\tablenotetext{b}{This is the coherent S/N of the FRB detected in the VLA image.} 
\tablenotetext{c}{Estimated using the Macquart-relation.}
\tablenotetext{d}{FRB20191228A shows a scattering tail of $6.1(6)$~ms. The deconvolved width is $2.3(6)$~ms. }
\tablenotetext{e}{The energies are derived assuming a flat spectrum for FRBs ($\alpha=0$) and zero k-correction.}
\tablenotetext{f}{In units of 12 + log[O/H].}
\tablenotetext{*}{These measurements are derived from spectroscopy. }
\tablenotetext{\dagger}{These properties are derived using \texttt{CIGALE} SED fitting.}
\tablenotetext{}{1$\sigma$ uncertainties are quoted for these measurements.} 
\end{deluxetable*}

\subsubsection{Spatial localization}
The images generated by the \rf\ search pipeline make several assumptions (coarse DM grid, non-optimal image size, simpler calibration algorithm, etc) during calibration and imaging. To address these, we used the raw, de-dispersed burst visibilities to form the burst image using CASA \citep{CASA}. We ran the full CASA pipeline to generate VLA calibration tables for this observation. We then used the CASA task \texttt{applycal} to apply those calibration and flagging table to the burst data. Observations of calibrator 3C147 were used to calibrate the flux density scale, bandpass and delays. The nearby source J0632$+$1022 was used to calibrate the complex gain fluctuations over time, by observing it every 20 minutes. We then used CASA tasks \texttt{tclean} and \texttt{imfit} to generate a radio image and fit an ellipse to the burst position. The initial burst position was found to be RA(J2000): 06h12m54.47s and Dec(J2000): +04d40$'$15.6$\arcsec$ with a statistical uncertainty of 0.01$\arcsec$ in RA and 0.01$\arcsec$ in Dec (see bottom panel of Fig.\,\ref{fig:realfast}).
\begin{figure*}
\begin{tabular}{ccc}
\includegraphics[trim=0 1 70 1,clip,scale=0.26]{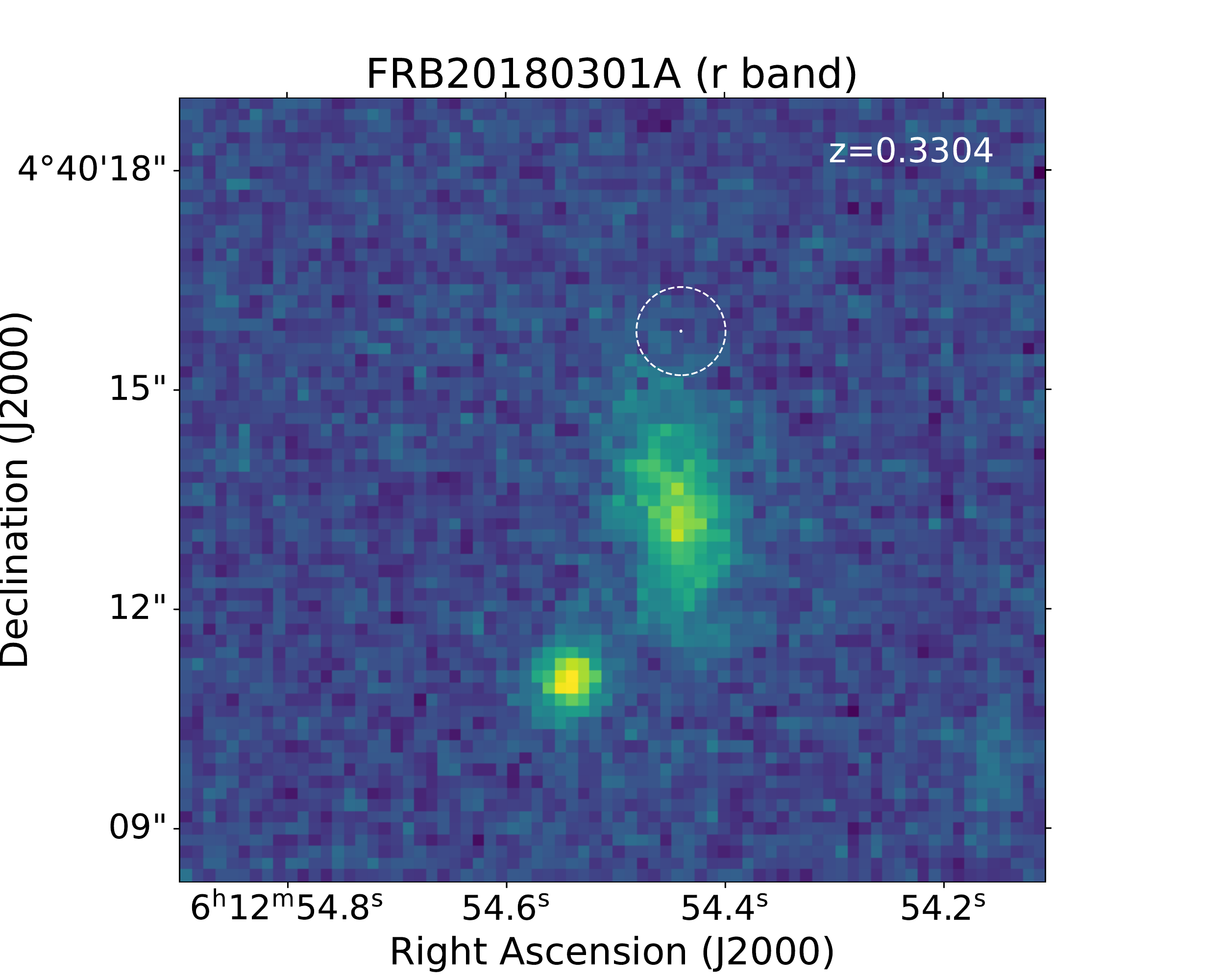}
\includegraphics[trim=1 1 70 1,clip,scale=0.26]{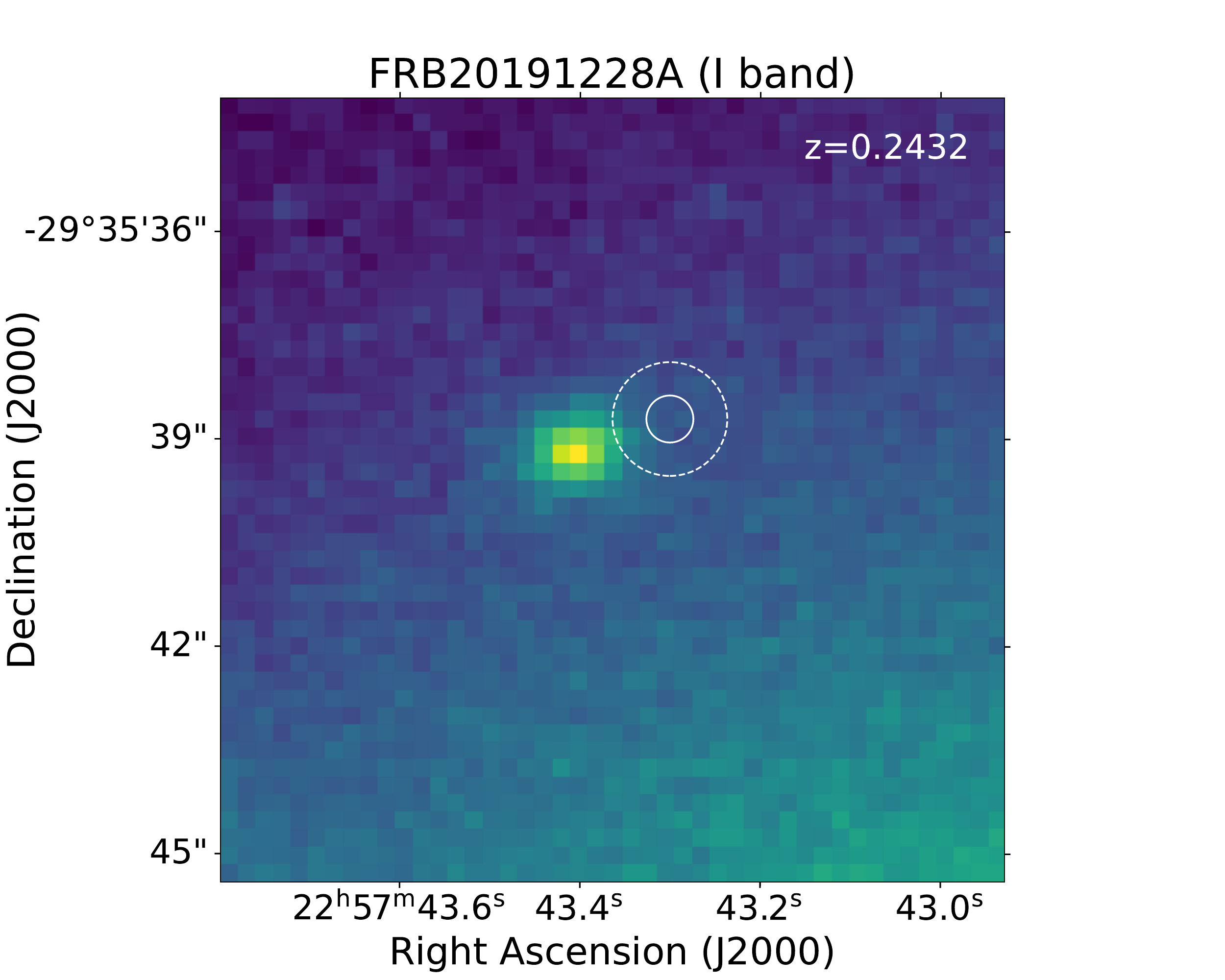}
\includegraphics[trim=1 1 1 1,clip,scale=0.26]{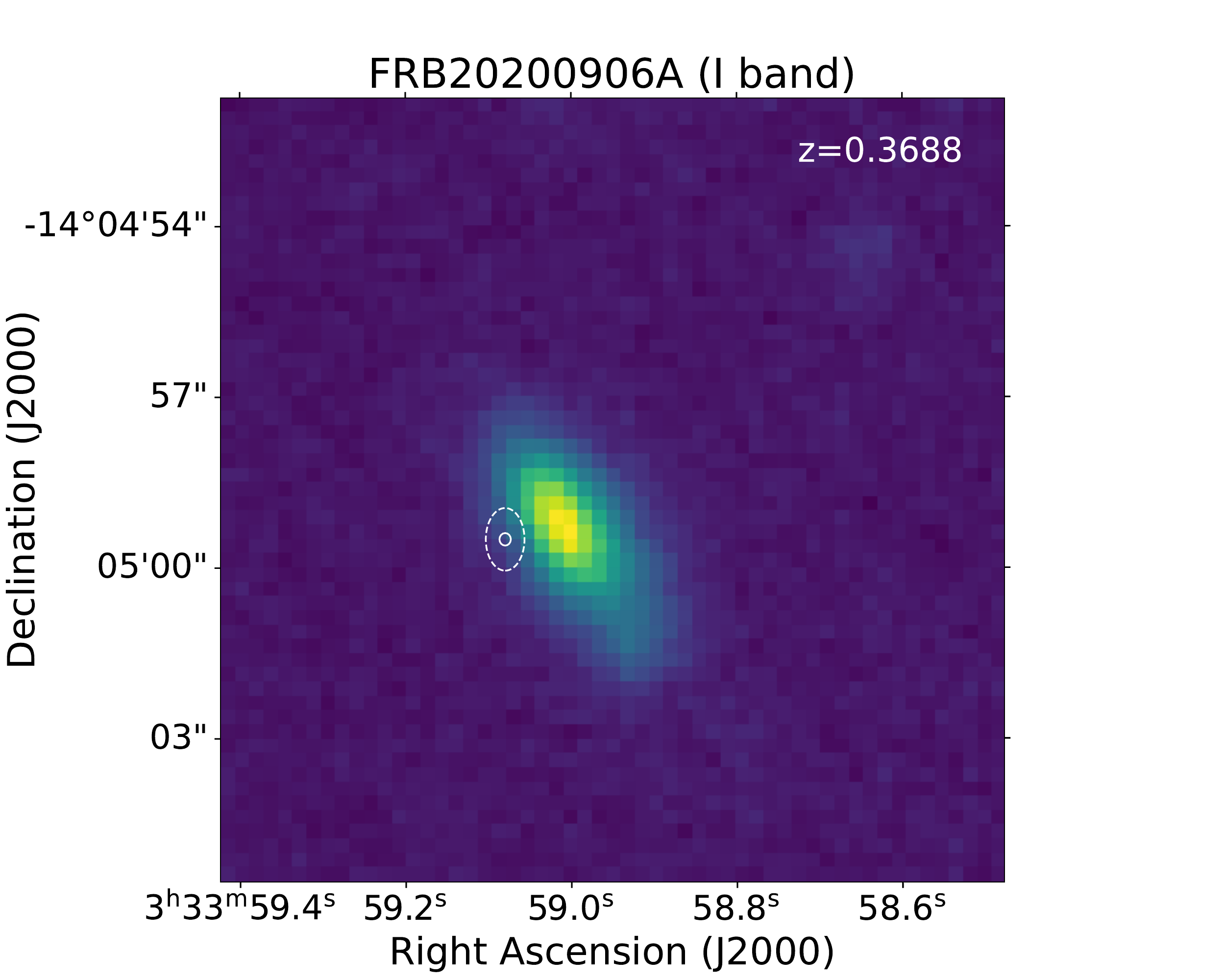}
\end{tabular}
\caption{$r$-band Gemini image of the host of repeating FRB20180301A and $I$-band FORS2 images of the hosts of FRB20191228A and FRB20200906A, overplotted with the positions of each FRB. The white solid and dashed circle/ellipse represents the $1\sigma$ statistical and systematic uncertainty respectively in the FRB position. }
\label{fig:host}
\end{figure*}

\subsubsection{VLA Radio Continuum Image}\label{sec:dri}
In addition to the fast-sampled data with an integration time on order of a few milliseconds generated by \rf, the VLA correlator creates slow-sampled data with an integration time of 5\,s. This slow sampled data will average out any fast-varying signal (like FRBs) but is suitable for finding any persistent radio source near the burst location. We made separate images for each epoch and combined images to search for any persistent or slowly varying emission from the FRB location.

The slow-sampled data was calibrated using the standard VLA calibration pipeline, followed by manual data inspection and further flagging. We used \texttt{tclean} to produce images. Manual flagging of the dataset was done to allow for much finer and in-depth RFI flagging than autoflagging alogrithms. The inputs of \texttt{tclean} used the gridder \texttt{wproject} with 128 wprojplanes. The deconvolver was set to \texttt{mtmfs} and \texttt{Briggs} weighting was used with a robustness parameter of 0. These parameters were chosen to help reduce noise from sources outside the primary beam as well as suppressing minor baseline-dependent errors. 
The full field of view was also imaged to allow cleaning on these bright sources, specifically sources $\sim42$ and $\sim30$
arcminutes away from the phase center.

Once \texttt{tclean} was run, the RMS near the center of the primary beam was found to be in the range $23-26\,\upmu$Jy~beam$^{-1}$ for all epochs. The image RMS is consistent with the combined effect of confusion noise ($18\,\upmu$Jy~beam$^{-1}$) and thermal noise ($16\,\upmu$Jy~beam$^{-1}$\ in each epoch). Combining the first four epochs and last four epochs produces images with RMS of $21\,\upmu$Jy~beam$^{-1}$, consistent with reduced thermal noise added to a fixed confusion noise. We find no persistent radio source above a 3$\sigma$ flux density of $63\,\upmu$Jy\,beam$^{-1}$ yielding a luminosity limit of $1.8 \times 10^{22}$~W~Hz$^{-1}$ at 1.5~GHz.

\begin{figure}
\centering
\includegraphics[scale=0.35]{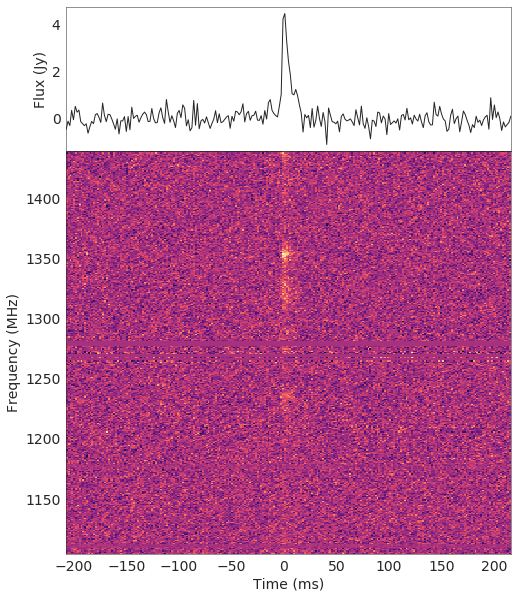}
\includegraphics[scale=0.35]{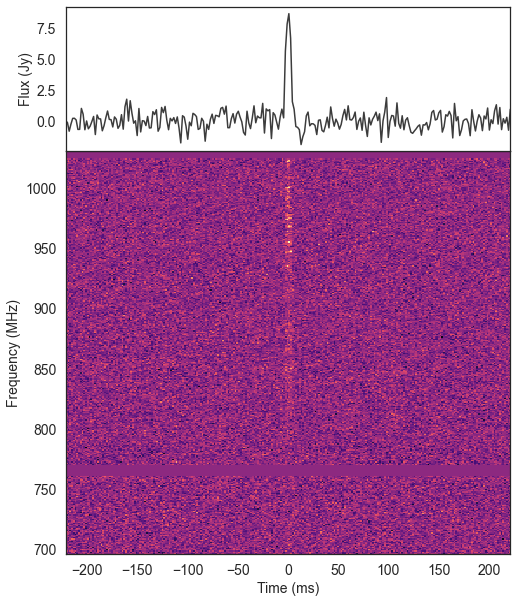}
\caption{Pulse profile and dynamic spectrum for FRB20191228A (top) and FRB20200906A (bottom). FRB signals are de-dispersed at the S/N maximised DM of 297.5\,pc\,cm$^{-3}$ and 577.8\,pc\,cm$^{-3}$ for FRB20191228A and FRB20200906A respectively.  }
\label{fig:waterfall}
\end{figure}

\subsubsection{Astrometry}
We tested our astrometric precision by associating radio
sources in the FRB-epoch VLA image with the Karl G. Jansky Very Large Array Sky Survey \citep[VLASS]{VLASS+20} Epoch 1 Quick Look catalog \citep{Gordon+20}. We re-imaged the FRB epoch, selecting an interval of 30\,s centered on the FRB for investigating astrometric accuracy. We ran the \texttt{PYSE} source finding package \citep{Pyse} on this 30\,s image and identified six radio sources above a significance level of $15\sigma$. All these sources had VLASS counterparts within a 2$\arcsec$ search radius. 
We measured the weighted mean and associated uncertainty of the offsets between these six sources in the 30\,s snapshot image and their VLASS catalog positions, measuring $156 \pm 330$\,mas and $-197 \pm 223$\,mas in RA and Dec respectively. However, the scatter in the measured offsets is clearly not due solely to the statistical (signal-to-noise limited) measurement uncertainty, with the $\chi^2$ of the weighted mean fit being 50.40 for 10 degrees of freedom. This is not unexpected, given the potential for systematic differences in centroid position due to, e.g., the different frequency and resolution of our 30\,s image vs the VLASS catalog.  Previous studies using ASKAP \citep{Day+21} have examined a large number of fields to estimate a scaling factor that can be applied to the weighted mean uncertainty to correct (on average) for these contributions, but the value of the scaling factor should depend on the telescope and reference catalogs used. Here, we instead take an unweighted mean of the six source offsets, which effectively assumes that the signal-to-noise limited contribution to each measured offset is small, and yields a more conservative estimation of the mean offset and uncertainty of $556\pm 619$\,mas and $-297\pm 603$\,mas in RA and Dec respectively.
We note that the astrometric accuracy of the VLASS Epoch 1 is well characterized, with a mean offset of the order $0.0\arcsec$ in RA and $0.15 \arcsec$ in Dec\footnote{\url{https://library.nrao.edu/public/memos/vla/vlass/VLASS_013.pdf}}. We corrected for this positional offset in the reference VLASS positions, and obtain a final offset correction of $0.75 \arcsec$ in RA and $-0.30 \arcsec$ in Dec. The final position of FRB20180301A after correcting for positional offsets along with the statistical and systematic uncertainties is  RA(J2000): 06h12m54.44s\,($\pm0.01\arcsec$$\pm0.62\arcsec$) and Dec(J2000): +04d40$'$15.8\,($\pm0.01\arcsec$$\pm0.6\arcsec$). 
These are also listed in Table \ref{tab:measurements}. 

We conducted additional tests on the FRB-epoch data to investigate the astrometric stability on $\sim$ minute scales across the entire observation duration, generating a set of images with spacing 3\,minutes. The mean offset varied by up to half an arcsecond between adjacent 3\,min images, consistent with our previous experience of L-band observations and implying a potential gradient in the observed positional offsets of up to $\sim$0.2 arcseconds/minute. We noted, however, that the average offset derived from the 30\,s image centered on FRB does not lie on an interpolation between those derived from adjacent 3\,min images, suggesting some structure on $<<3$~min timescales. We note that any remaining variability on $<30$\,s timescales could lead to a small, uncorrected positional bias. 

\subsubsection{Optical follow-up}
In archival Pan-STARRS images we identify a faint galaxy, PSO~J093.2268$+$04.6703, as the putative host of FRB20180301A. We acquired deep (0.5 -- 0.7 hrs in each band) follow-up imaging of the FRB field with the Nordic Optical Telescope (NOT) using the Sloan filters $u,g,r,i,$ and $z$ on nights between UT 2021 October 26 to December 14, followed by additional deep ($36 \times 100$\,s) Gemini $r$-band observations on UT 2021 January 25. The Gemini image was astrometrically calibrated to match the Gaia DR2 catalog \citep{GaiaDR2, GaiaDR2_astro}, with a relative accuracy of $0.2\arcsec$. We note that there is an additional nearby source to the south of the identified host galaxy in the field of FRB20180301A. The morphology of this source is consistent with being point-like, so we classify it as a Galactic foreground star. 

Observations of the host with the Keck/DEIMOS spectrograph on UT 2020 September 17 determined the galaxy's redshift $z= 0.3304$ based on nebular emission lines such as [N~{\sc ii}] and H$\alpha$. The measured redshift was found to be consistent with the redshift range $(z=0.13-0.35)$ estimated by extragalactic DM in \cite{Luo+20Nat}. We also performed NIR observations of the galaxy in $JHK-$bands using the MMT telescope on UT 2021 February 27 \& 28. The photometric measurements derived from MMT imaging, along with those from NOT, were used by CIGALE \citep{cigale} for SED fitting to estimate the stellar mass of the galaxy (see Appendix for details). The deep $r-$band Gemini imaging was used for the probabilistic host galaxy association analysis. We find a Probabilistic Association of Transients to their Hosts (PATH) \citep{PATH} probability of 99.9\% for the association of FRB20180301A with the given galaxy. The host is presented in Fig.\,\ref{fig:host} and properties are listed in Table\,\ref{tab:measurements}.

\section{ASKAP localization of apparently non-repeating FRBs}
\label{sec:CRAFT_FRBs}
Here we report the discoveries of the single bursts from two new FRB sources, FRBs\,20191228A and 20200906A, discovered in the CRAFT incoherent-summed (ICS) searches with ASKAP \cite[][]{2019ascl.soft06003B}. 
The burst pulse profiles and dynamic spectra are shown in Figure \ref{fig:waterfall}.  

\subsection{FRB20191228A}
FRB20191228A was detected on 2019 December 28 at UT 09:16:16.44440 during CRAFT observations conducted with ASKAP, with 28 antennas in a 336~MHz band centered on 1271.5~MHz. The burst had a detection S/N of 23 at a $\rm DM=297.5(5)$\,pc\,cm$^{-3}$. The real-time detection in online incoherent-sum data triggered a download of 3.1\,s of voltages around the FRB. The offline correlation of voltages across 28 antennas and their interferometric analysis led to an initial localization of FRB20191228A.

On 2020 January 31, we used the Australia Telescope Compact Array (ATCA) in the frequency range of $1.1-3.1$\,GHz to perform astrometric observations of background sources near the position of FRB20191228A, to refine the position of the burst. In addition to four target background sources in the field of FRB20191228A (J2258$-$2929, J2255$-$2937, J2258$-$2955 and J2259$-$2957), we observed three gain calibrators (PKS2254$-$367, PKS2337$-$334, and PKS2255$-$282). For each target background source, we obtained three position estimates -- one for each gain calibrator, by applying phase calibration solutions derived from that calibrator. We observed a scatter of $\sim$80\,mas in RA and $\sim200$\,mas in Dec in the position of the background sources depending on which gain calibrator was used. To account for this systematic uncertainty to the ATCA positions, for each background source we took a weighted mean of the three positions to estimate the probable location, and added the average residual offset from this mean position in quadrature to our estimate of the absolute positional uncertainty for the source.

The positions of target sources obtained from the ATCA radio image were compared with those obtained from the 3.1\,s ASKAP data, following the method described in \cite{Day2020}, to astrometrically register the FRB image frame to that of the International Celestial Reference Frame \citep[ICRF3,][]{ICRF3_2018AGUFM.G42A..01G} and estimate the accuracy of this registration. J2258$-$2929 and J2259$-$2957 were excluded from the field source comparison due to low S/N in the ASKAP image for the latter, and the former being resolved.

Finally, we computed a weighted mean systematic image frame offset using the method described in \citet{Day+21}. We found positional offsets of $0.410 \pm 0.830\arcsec$ and $-0.856 \pm 0.823\arcsec$ in RA and Dec., respectively. After accounting for astrometric shifts in the image frame as well as statistical and systematic errors, the final FRB position is RA(J2000): 22h57m43.30s\,($\pm0.34\arcsec$$\pm0.83\arcsec$) and Dec(J2000): $-$29d35$'$38.7\,($\pm0.3\arcsec$$\pm0.8\arcsec$).

\subsubsection{Follow-up observations} \label{optical}
The burst position is 1.8\,arcmin from the first-magnitude star Fomalhaut ($\alpha$\,PsA), which severely complicated identification and characterisation of the optical counterpart. 

Deep imaging with FORS2 instrument at the European Southern Observatory's (ESO) Very Large Telescope (VLT) was performed in the $g$-band ($5 \times 90$\,s) and $I$-band ($20 \times 100$\,s) on 2020 September 21 UT, while using the movable slitlets normally employed for multi-object spectroscopy to completely mask the lower of the 2 CCDs to protect the detector from saturation by Fomalhaut. The seeing ranged between $0.6\arcsec-0.9\arcsec$ during the observations.

The individual frames were bias-subtracted and flatfielded using the ESOReflex package \citep{ESOReflex}. To overcome the glare produced by Fomalhaut, a two-dimensional polynomial was fit to a patch of sky centered on the burst position in each individual frame, after masking sources. This model was then subtracted from the patch. The glare-subtracted images were coadded with the \texttt{Montage} package \citep{Berriman2017}. A candidate host galaxy was identified with centroid approximately $1.5\arcsec$ from the burst coordinates with a 100\% PATH association probability (see Fig. \ref{fig:host}). Photometry of the identified galaxy was performed with \texttt{Source-Extractor} \citep{SExtractor} using a circular aperture with a $5\arcsec$ diameter. The effect of the glare subtraction on the photometry was investigated, and the uncertainty introduced by the procedure quantified, using injected synthetic sources of known magnitude (making use of \texttt{Astropy}, \citealp{astropy1}).

On UT 2020 November 08, the Keck/LRIS spectrograph was used to perform spectroscopic observations in order to determine the galaxy redshift. The nebular emissions such as H$\alpha$, [O~{\sc iii}] established the redshift of the host to be $z=0.243$.   

We also performed observations of the host of FRB20191228A using the ATCA at center frequencies of 5.5 and 7.5\,GHz to search for a compact and persistent radio emission. We did not detect any radio emission above 22\,$\upmu$Jy beam$^{-1}$ (3$\sigma$) constraining the luminosity of the source to be $< 3.4 \times 10^{21}$~W~Hz$^{-1}$ at 6.5\,GHz.  

\subsection{FRB20200906A}
FRB20200906A was discovered in the incoherent sum of 7 ASKAP antennas on 2020 September 06 UT 21:40:50.923 during CRAFT observations at 864.5~MHz. The burst had an optimal S/N of 19.2 at a DM of 577.8(2)\,pc\,cm$^{-3}$ in the low time resolution search data stream. Voltages spanning 3.1\,s around the FRB were downloaded, cross-correlated and imaged offline to obtain a preliminary position of the burst. The astrometric registration was performed by comparing background sources in the 3.1\,s ASKAP image with their NRAO VLA Sky Survey (NVSS) radio source catalog \citep{NVSS} counterparts. We cross-matched seven sources and computed a weighted mean systematic image frame offset using the method described in \citet{Day+21}. We found positional offsets of $2.05 \pm 0.34\arcsec$ and $0.51 \pm 0.55\arcsec$ in RA and Dec., respectively. The final burst position after correcting for the astrometric shifts in the image frame is RA(J2000): 03h33m59.08s\,($\pm0.10\arcsec$$\pm0.34\arcsec$) and Dec(J2000): $-$14d$04'59.5$\,($\pm0.1\arcsec$$\pm0.6\arcsec$), where statistical and systematic uncertainties in RA and Dec are quoted respectively. 

\subsubsection{Follow-up observations}
A candidate host galaxy for FRB20200906A was identified in the DES, Pan-STARRS and AllWISE database as DES~J033358.99$-$140459.2, PSO~J033358.994$-$140459.287 and J033358.99$-$140459.1 respectively, with 100\% PATH association probability (see Fig.\,\ref{fig:host}). On UT 2020 December 20 \& 22, we performed follow-up observations using VLT/FORS2 in $g$- and $I$-band following a similar strategy as for FRB20191228A (Sec. \ref{optical}) but without the need to mask the lower CCD. On UT 2020 September 17, we also triggered Keck/DEIMOS for spectroscopic observations and found the redshift of the putative host to be $z=0.3688$. The photometric measurements from Pan-STARRS and those from the VLT were used to model the SED of the host galaxy. The derived properties are presented in Table \ref{tab:measurements}. 

We triggered the VLA (project code: VLA/20A-157) on UT 2020 September 22 to observe the host galaxy of FRB20200906A in the frequency range $4-8$\,GHz. We found no radio emission from the host galaxy above a $3\sigma$ flux density of 12\,$\upmu$Jy~beam$^{-1}$, implying an upper-limit on the source's luminosity to be $< 4.3 \times 10^{21}$ W\,Hz$^{-1}$ at 6\,GHz. 
\begin{figure}
\centering
\includegraphics[scale=0.4]{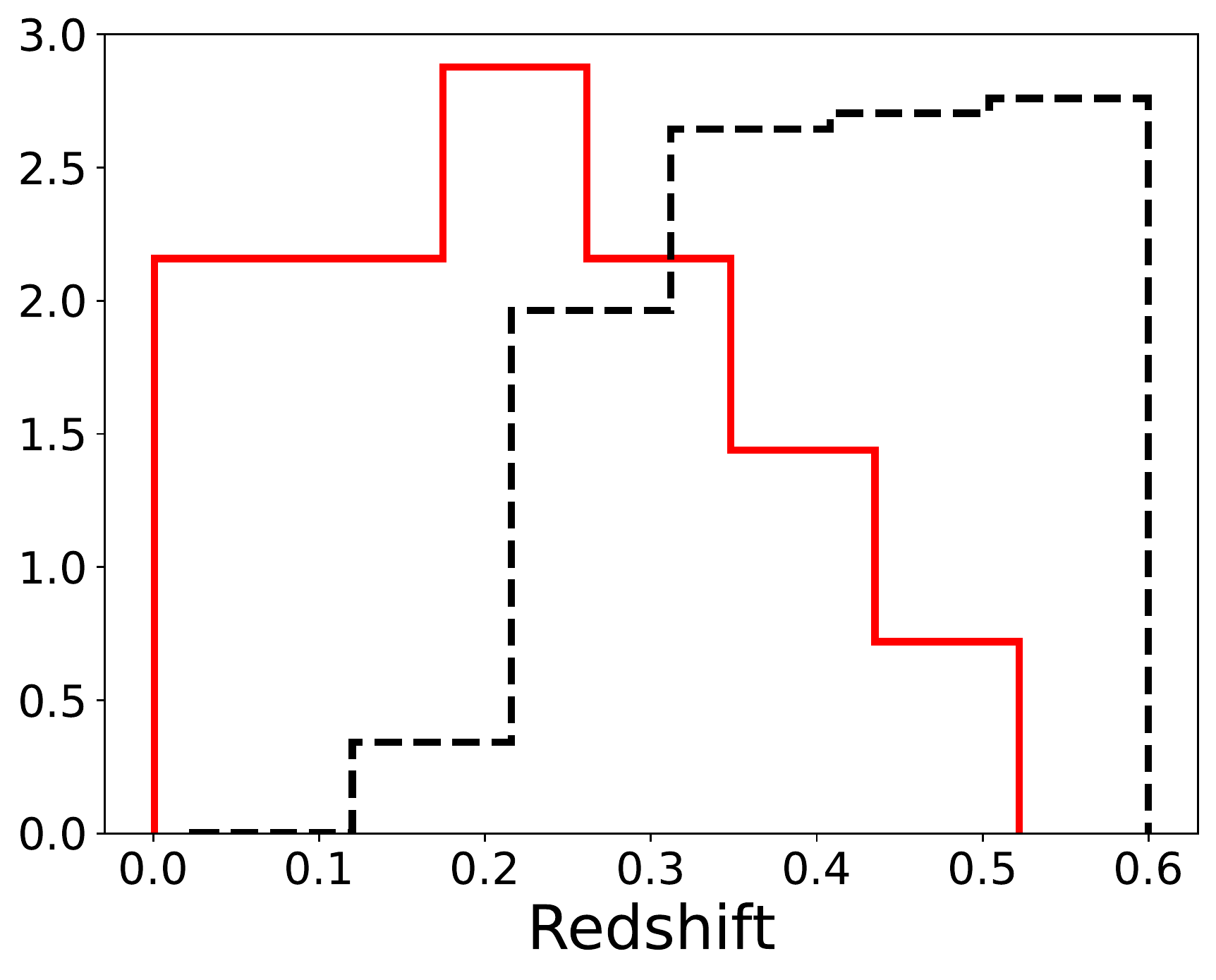}
\caption{Redshift distribution of the FRB hosts (red) and PRIMUS sample of field galaxies (dashed black) where the the area under the histogram integrates to 1. The PRIMUS sample lack good stellar mass and SFR measurements of field galaxies for $z <0.2$.}
\label{fig:redshift}
\end{figure}

\begin{figure*}[ht]
\begin{tabular}{cc}
\includegraphics[scale=0.42]{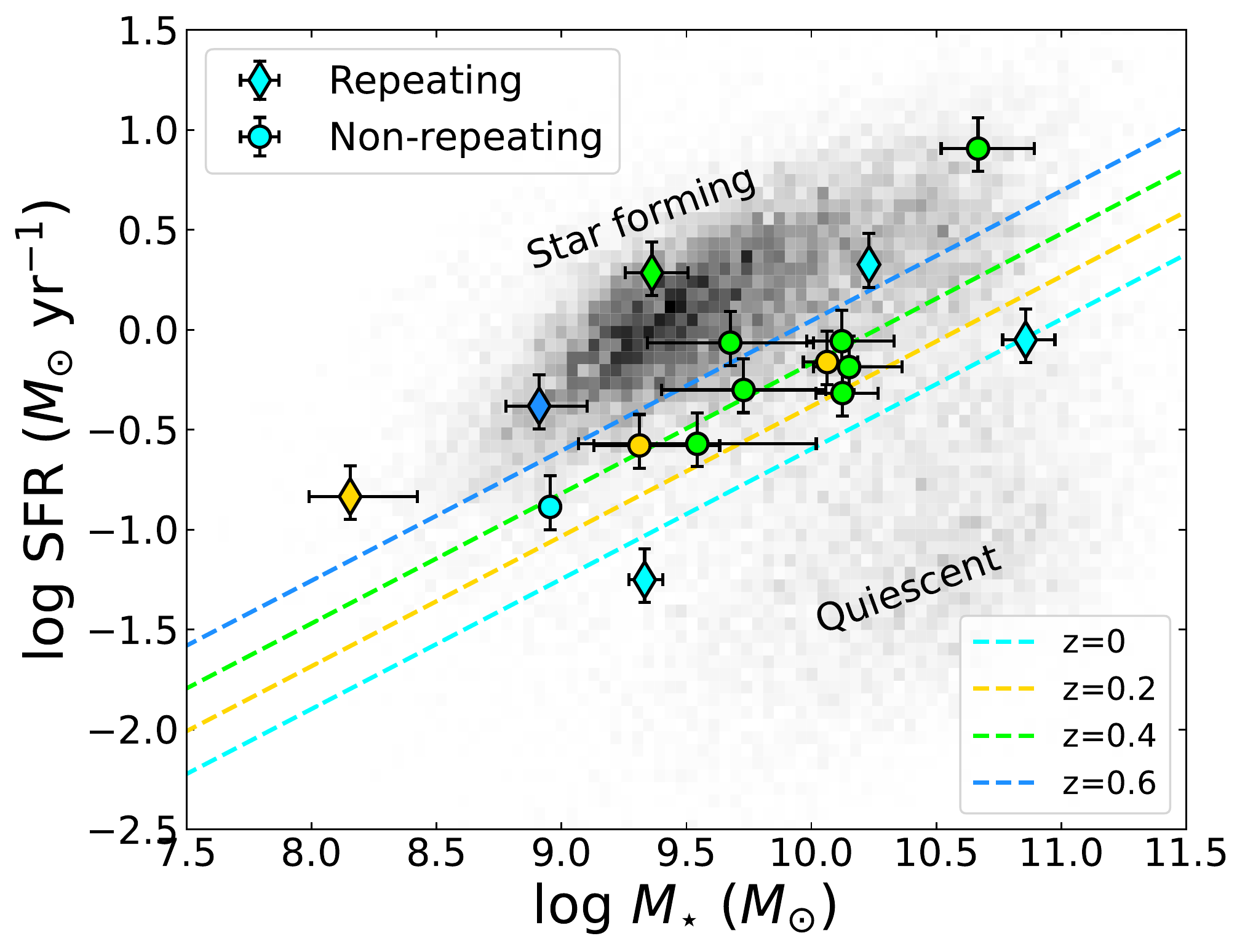}
\includegraphics[scale=0.42]{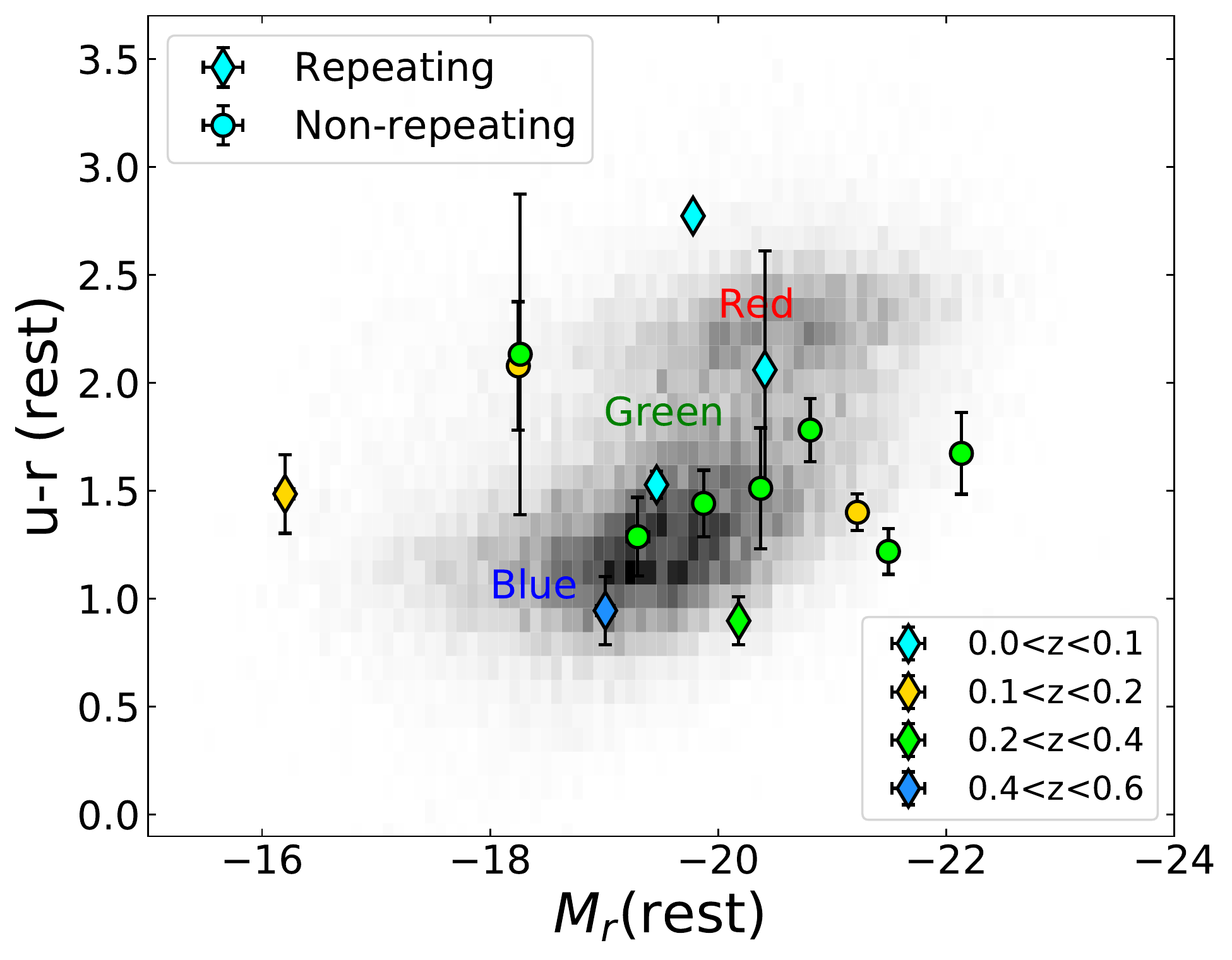}

\end{tabular}
\caption{Left: Star-formation rate and stellar mass distributions of the host galaxies of 6 repeating and 10 one-off FRBs compared against a population of galaxies at $z<0.6$ taken from the PRIMUS survey \citep{PRIMUS}. Diamond and circle symbols represent the hosts of repeating and non-repeating FRBs respectively. The boundary separating the star-forming and quiescent galaxies and its evolution with redshift \citep{PRIMUS} is presented by coloured dashed lines. Right: Rest-frame color-magnitude diagram of the host galaxies of a sample of 6 repeating and 9 one-off FRBs compared to the population of PRIMUS galaxies at $z<0.6$. FRB20171020A is not included due to lack of $u-r$ measurement. We also note that the observed magnitudes for FRB20201124A and FRB20200120E are approximated as rest-frame magnitudes because of their low redshifts and thus, small k-correction. For both plots, the FRB host data is color-coded and divided into four redshift bins of $<0.0<z<0.1$ (cyan), $<0.1<z<0.2$ (yellow), $<0.2<z<0.4$ (green) and $<0.4<z<0.6$ (blue). 
}
\label{fig:global_prop}
\end{figure*}

\begin{figure}
\centering
\includegraphics[scale=0.4]{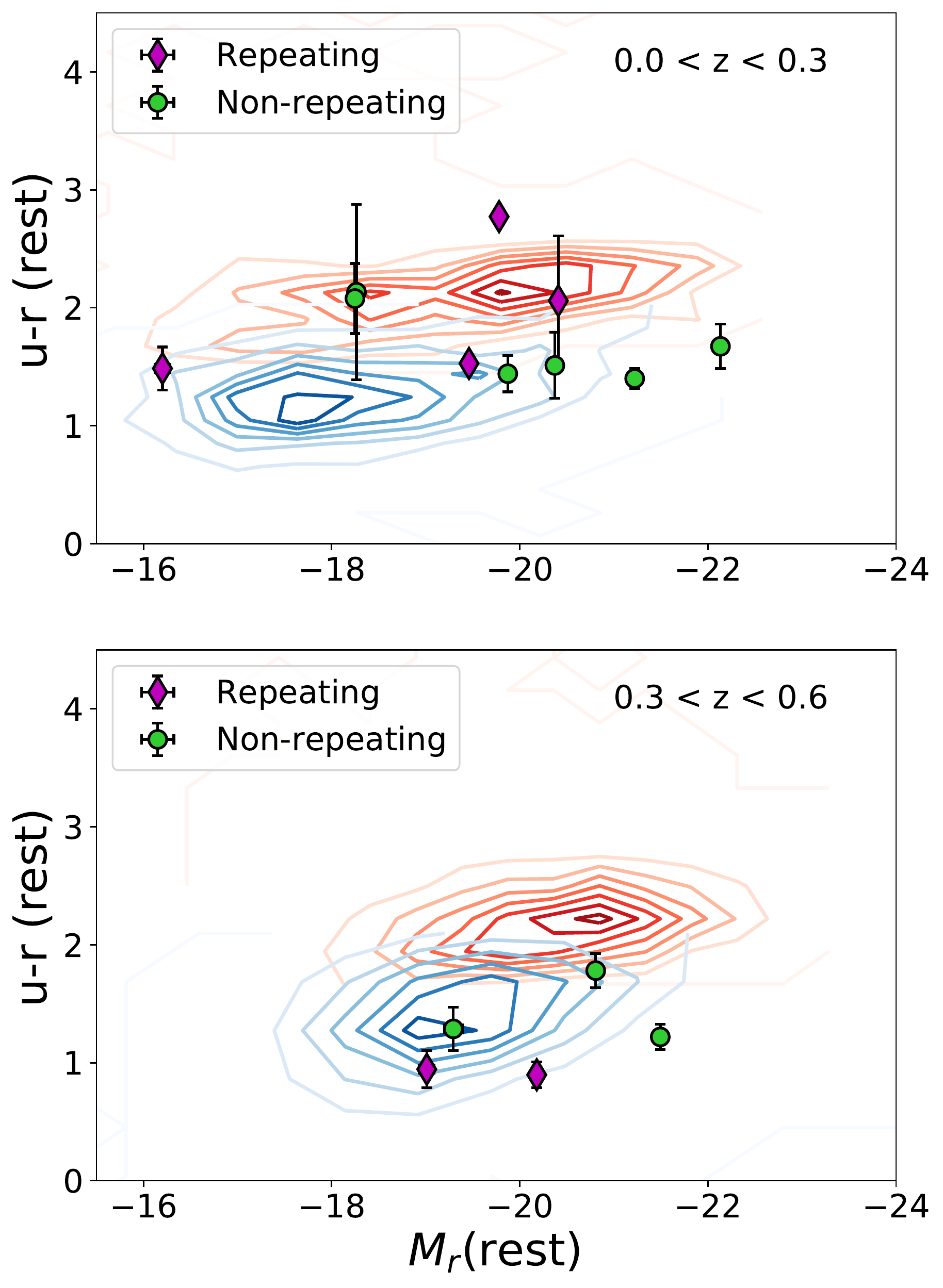}
\caption{Approximation of the stellar density in blue and red galaxies as presented by blue and red contours respectively and derived from the PRIMUS galaxies in the color-magnitude diagram for the two redshift bins. The shading from white to respective color represent outer to inner contour levels. The contour levels are set same for blue and red galaxies. The bright red region shows that most of the stars in the local Universe are concentrated in massive red galaxies. Green and magenta data points are the sample of one-off and repeating FRB hosts respectively. We observe that FRBs are distributed in normal blue galaxies.}
\label{fig:Mass_weighted}
\end{figure}

\begin{figure}
\centering
\includegraphics[scale=0.33]{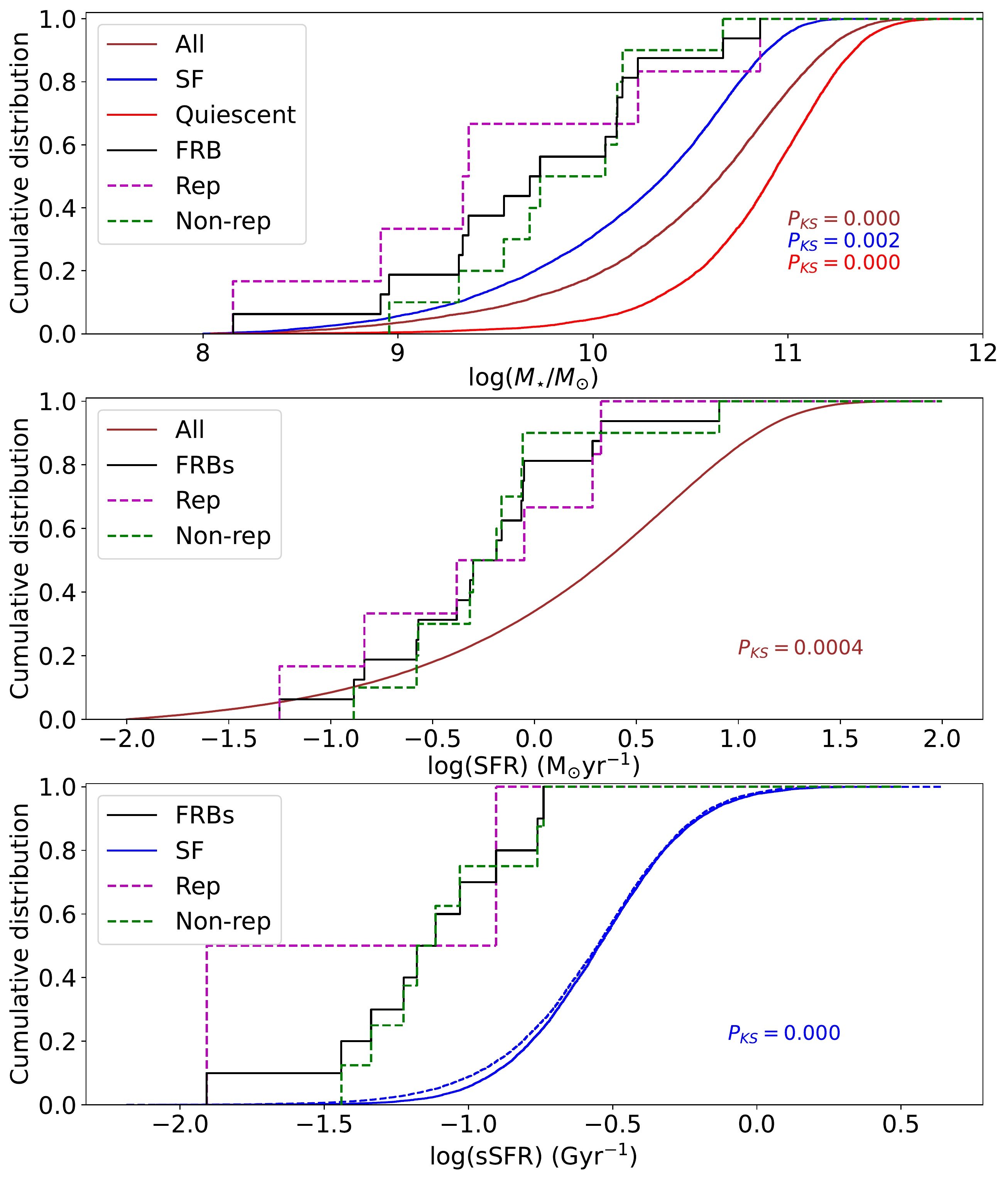}
\caption{Top: Comparison of FRB cumulative mass distribution with mass-weighted stellar mass distribution for all field galaxies (brown), star-forming galaxies (blue) and quiescent galaxies (red). Middle: Comparison of FRB host cumulative SFR distributions (black) with the star formation rate volume density distribution function for all field galaxies in the local Universe (brown). Bottom: Comparison of sSFRs for FRBs in the mass range $9.5 < {\mathrm log(M)} < 11.5$ with the log-normal (solid) and double exponential functions (dashed) for sSFRs of main-sequence star-forming galaxies. }
\label{fig:frb_functions}
\end{figure}
\section{Comparing FRB host galaxies to the underlying field galaxy population}
\label{sec:FRBs_galaxies}
In this section, we place the overall population of FRB hosts in context with the field galaxy population, expanding on our earlier work \citep{Bhandari_2020a,Heintz+20}. 
\subsection{Sample selection}
We use the data published in \citet{Heintz+20}, \citet{Mannings+20} and measurements derived for the hosts in this work. We also update the measurements published for a subset of FRB hosts (see Appendix). For FRB20171020A, FRB20201124A and FRB20200120E, we use the data published in \citet{Mahony2018}, \citet{Fong+21} and \citet{Bhardwaj+21} respectively. 

We considered only the FRB host galaxies for which the PATH posterior probability is greater than 90\% (see last column of Table\,\ref{tab:results}). 
This includes all 3 of the new FRBs presented here which have
posterior probability $P(O|x) > 0.999$. The host of FRB20190614D \citep{Law2020}, FRB20181112A \citep{X+19} and FRB20190523A \citep{Ravi+19a} have low probabilities of $P(O|x)$ of $0.58, 0.67$ and $0.82$ respectively and, are therefore excluded.  
Also, FRB20191001A has two nearby candidate hosts at a common redshift \citep{Bhandari_2020b}, that yield PATH posterior probabilities $P(O|x) = 0.6$ and 0.4. 
We proceeded by adopting the galaxy with smaller angular separation from the FRB as the host which is
akin to adopting a stronger prior on the projected
offset than adopted by \citet{PATH}.
Future associations will refine this assumption. We also included the host of FRB20200120E \citep{Bhardwaj+21,Kirsten+21} and the 
candidate host for FRB20171020A, because of their low probability of chance association (North-Hickey et al. in prep). Thus, we obtained a sample of 16 confident
host associations for our analysis, of which 10 are hosts of apparently non-repeating FRBs and six host repeating FRBs (see Table\,\ref{tab:updated_properties}). 

The spectroscopy and photometry of all FRB hosts are analyzed with the \texttt{pPXF} \citep{ppxf} and \texttt{CIGALE} \citep{cigale} software packages by fitting a set of stellar population models and star formation history to the spectra and spectral energy distribution (SED) respectively (See \citet{Heintz+20} for details). The star-formation rates (SFR) are derived from the dust-corrected H$\alpha$ line flux measurements \citep{Kennicutt98}, adopting an initial mass function (IMF) from \citet{Chabrier03}. 

We used the PRism MUlti-object Survey (PRIMUS) data as an underlying sample of field galaxies \citep{PRIMUS}. We restricted the redshift range to be $z < 0.6$ to match the redshifts of the FRB hosts (see Fig.\,\ref{fig:redshift}). We used K-correct synthesized restframe SDSS absolute magnitude of $\sim 108,000$ galaxies and stellar masses/SFRs of $\sim 31,200$ galaxies. We note that the PRIMUS sample lack good stellar mass and SFR measurements of field galaxies for $z<0.2$. All stellar masses and SFRs assume a universal \citep{Chabrier03} IMF and are derived using SED-modeling code \texttt{iSEDfit}, designed to extract the physical
properties of galaxies \citep{PRIMUS}.

\subsection{Comparison with underlying population}
Figure\ \ref{fig:global_prop} shows our comparison of the color$-$magnitude and SFR$-$M$_*$ distributions of the FRB host galaxies with those of the general population of galaxies at redshift $<0.6$. The left panel of Fig.\,\ref{fig:global_prop} presents the current SFR as a function of M$_*$. The hosts of repeating and one-off bursts are distinguished by symbol shape and the color represents the four redshift bins. We also present the redshift evolution of the boundary separating the star-forming and quiescent galaxies \citep{PRIMUS}.
Most of the FRB hosts lie in or around the star-forming cloud of galaxies, but are offset from the star-forming main sequence for galaxies with similar stellar masses. 
We caution here that since a significant fraction of the FRB hosts show LINER-like emission (see Sec.\,\ref{section:BPT}), the derived SFR should in these cases only be considered as upper limits since the total line emission may not reflect solely that of star formation. This would further offset the FRB hosts from the star-forming main-sequence.  

The right panel of Fig.\,\ref{fig:global_prop} shows a color magnitude diagram and provides information about the overall stellar populations in these galaxies. The late-type galaxies with ongoing star formation and therefore, young stellar populations lie in the `blue cloud', i.e., blue galaxies \citep{Strateva2001}, while massive early-type galaxies live in the `red and dead' zone characterized by very low star formation and hence older stellar populations, i.e., red galaxies. The host galaxies of the FRBs appear to lie on the luminous side of the absolute magnitude distribution, mainly near the `blue cloud' and `green valley' region, where galaxies are expected to be transitioning between star-forming and quiescent systems \citep{Martin2007b}. We observe a dearth of red galaxies in our current sample of FRB host galaxies.

\subsection{Do FRB hosts track stellar mass and star formation rates?}
In Fig.\,\ref{fig:Mass_weighted} we again present the color-magnitude diagram, but with the background PRIMUS galaxy sample weighted by their stellar masses and divided into redshift bins of $0.0<z<0.3$ and $0.3<z<0.6$. This is a good approximation of where most stars are in the local Universe. We show that the majority of FRB hosts do not trace massive red galaxies and also do not seem to align with the color-magnitude diagram space of the peak of stellar mass-weighted
blue galaxies which tend to redder color, particularly for the low-redshift bin. To quantify this trend, we examined the null hypothesis that FRB hosts (both repeating and non-repeating population) track stellar mass.
The galaxy stellar mass function (GSMF) of low-$z$ galaxies, $\Phi(M_{*})\Delta M$, was weighted by stellar mass and compared to the observed FRB host mass distribution. We used the double Schechter function to model GSMF for all, star-forming and passive galaxies in the redshift range $0.2 < z < 0.5$ in the COSMOS field \citep{Davidzon+17}. Our KS tests comparing the cumulative mass distributions of all FRB hosts, together and separately for the two FRB populations, with the mass-weighted stellar mass distribution of field galaxies yield a p-value $P_{KS} < 0.05$ (see Table.\,\ref{tab:pvalues_frbfunction}). Therefore, we reject the null hypothesis that FRB hosts directly track stellar mass with more than 95\% confidence, consistent with the findings of \citet{Heintz+20}. We note that the redshift range of FRB hosts is broader than the range used for the stellar mass-weighted mass function. We replicated the above analysis using a subset of FRBs in the range $0.2 <z< 0.5$ to investigate if the differences in the redshift had an effect on our conclusions. We obtained identical results, indicating that the effect of redshift evolution is not substantial.

Furthermore, we tested the null hypothesis that FRB hosts track star formation rates. We used the star formation rate distribution function derived from the UV and IR luminosity Schechter functions for the local Universe using GALEX data \citep{Bothwell+11}. We then computed the star formation rate volume density distribution function, which is given by $\Phi(\Psi) \times \Psi$, where $\Psi$ is the star formation rate in M$_{\odot}$~yr$^{-1}$. The p-values for the KS test performed between the cumulative star formation rate volume density distribution function and the star formation rates of FRB hosts (both repeating and non-repeating population) in our sample are presented in Table.\,\ref{tab:pvalues_frbfunction}. We find the p-values for all and non-repeating subset to be $<10^{-3}$. [Considering the SFRs as upper limits due to possible LINER emission contamination (see Sect\,\ref{section:BPT}) would make the discrepancy even more significant]. However, when comparing with the repeating host population, we find a p-value slightly higher than our significance level. As a result, while we reject the null hypothesis with greater than 95\% confidence for all FRB hosts, we are unable to reject it with the same level of confidence for the repeating host population. We note a possible caveat of comparing the SFRs derived using the UV and IR luminosity for the GALEX sample with those derived using H$\alpha$ emission-line luminosity for the FRB host sample. Nevertheless, \citet{Lee2019} showed a coarse agreement between the FUV and H$\alpha$ SFRs, where the SFRs agree to within about a factor of two for the majority of galaxies with SFR~$\geq 0.01$\,M$_{\odot}$\,yr$^{-1}$.


Finally, we investigated the specific star formation rates (sSFRs) of FRB hosts in our sample with that of sSFR functions derived from the star-forming sample of galaxies defined by a main sequence using a color–color selection in COSMOS and GOODS survey in the redshift range $0.2-0.4$ \citep{Ilbert+15}. The sSFR function can be modelled as a log-normal or a double exponential profile for a given stellar mass bin range (Eq.\,2 and 3 of \citet{Ilbert+15}). We computed the function over four mass bins ranging from $9.5 < \mathrm{log(M)} < 11.5$ and combined them together. The weighted cumulative sum of this sSFR function is compared to the cumulative distribution of sSFRs of FRB hosts in the same mass range. The results are presented in Table \,\ref{tab:pvalues_frbfunction}. We observed that FRB hosts do not follow the sSFRs of star-forming galaxies and thus reject the null hypothesis with more than 95\% confidence. 

Thus, we conclude that FRB hosts have lower M$_*$, SFR and sSFR
than randomly selected field galaxies weighted by M$_*$, SFR or sSFR. 
\definecolor{LightCyan}{rgb}{0.88,1,1}
\definecolor{Gray}{gray}{0.9}
\begin{deluxetable}{ccccccccc}
\tablewidth{0pc}
\tablecaption{P-values for two-sample 1D-KS for comparing the mass-weighted stellar mass distributions (of all, star-forming (SF) and quiescent (Q) sample), star formation rate volume density distribution and the double exponential function for sSFRs of field galaxies with FRB host (repeating and non-repeating) cumulative mass, SFR and sSFR distribution.  \label{tab:pvalues_frbfunction}}
\tabletypesize{\footnotesize}
\tablehead{\colhead{FRB Type} & \multicolumn{3}{c}{Stellar mass} & \colhead{SFR}  
& \colhead{sSFR} 
\\
\colhead{} & \colhead{All} & \colhead{SF} & \colhead{Q} & \colhead{All} & \colhead{SF} \\
} 
\startdata 
All & 4e-6 & 0.002 & 4e-11 & 4e-4 & 5e-6 \\
Rep  & 0.014 & 0.029 & 6e-4 & 0.056 & 0.037 \\
Non-rep  & 7e-5 & 0.005 & 4e-8 & 0.001 & 7e-6 \\
\hline
\enddata 
\end{deluxetable} 

\begin{figure}
\includegraphics[scale=0.37]{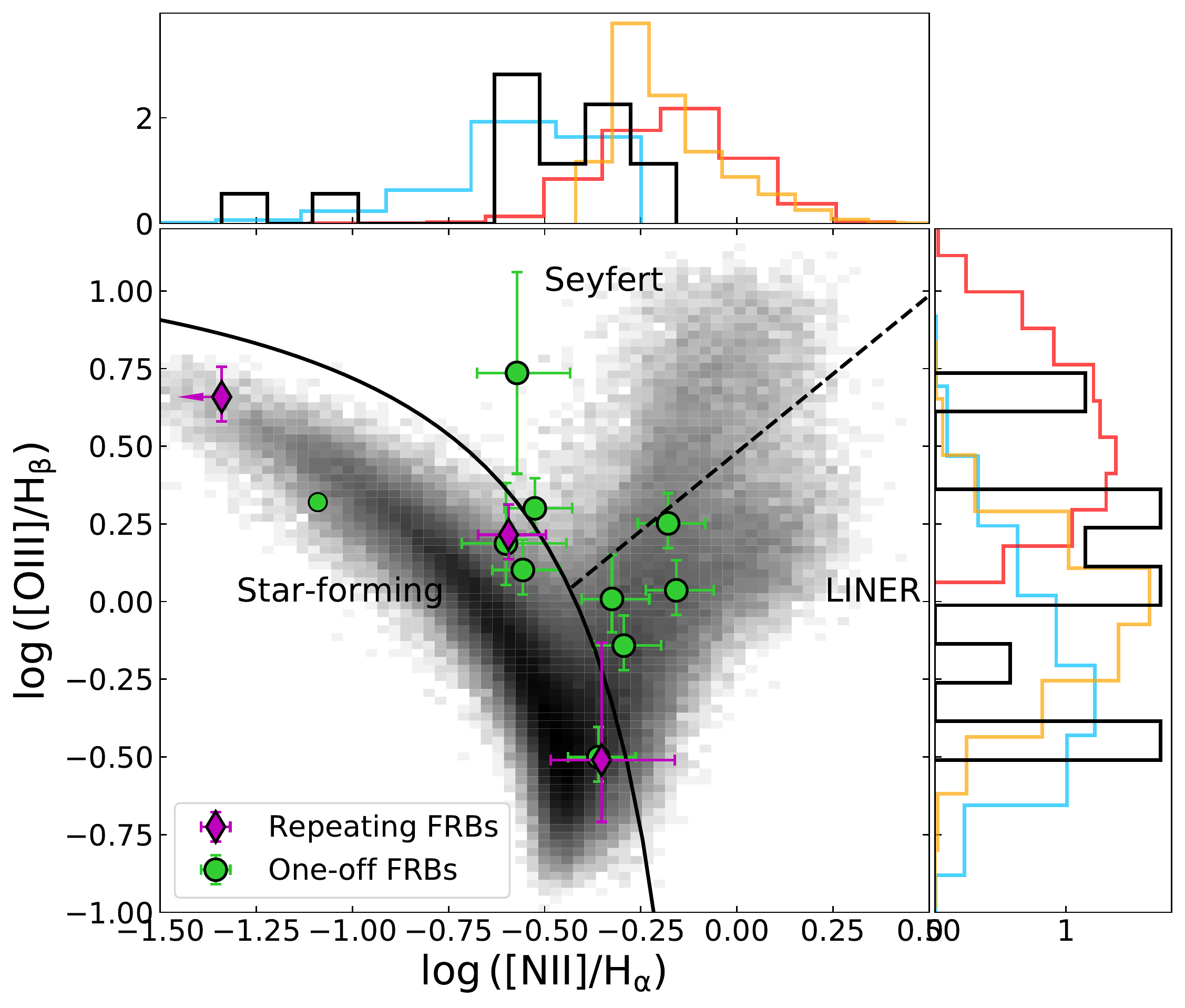} 
\caption{BPT classification diagram for FRB hosts The gray-scale background shows the density distribution of SDSS galaxies with redshifts $0.02 < z < 0.4$. The dashed and dotted black lines represent the demarcation line between SF galaxies and AGNs \citep{kauffmann03} and AGN and LINERs \citep{fernandes10}, respectively. The majority of FRB hosts are SF and LINER galaxies. Top and sideways inset present the histograms of the ratio of [N~{\sc ii}]/H$\alpha$ and [O~{\sc iii}]/H$\beta$ respectively for FRBs (black), AGNs (red), SF (light blue) and LINER (orange) galaxies. We note that a smaller sample of FRBs is shown in this plot, due to either lack of available spectral line data or the non-detection of required emission lines. }
\label{fig:bpt}
\end{figure}
\subsection{BPT diagram}
\label{section:BPT}
To identify the dominant source of ionization in FRB host galaxies, we plot their nebular emission line ratios namely, [N~{\sc ii}]/H$\alpha$ and [O~{\sc iii}]/H$\beta$ in a Baldwin-Phillips-Terlevich (BPT) diagram \citep{BPT81} in Fig.\,\ref{fig:bpt}. This diagram can also be used to distinguish between star-forming (SF) galaxies, low-ionization nuclear emission-line region (LINER) galaxies, and Active Galactic Nuclei (AGN) (see \citet{kewley01} for more details). For comparison, we show the distribution of nearby $(0.02 < z < 0.4)$ emission-line ($>5\sigma$ significance) galaxies from the Sloan Digital Sky Survey (SDSS) including the standard demarcation lines between SF, AGN, and LINER galaxies \citep{kauffmann03,fernandes10}. We performed a 2D KS-test to compare the host population of FRBs with that of underlying SF, AGN and LINER galaxies and found that they are not drawn from a specific class ($P_{\rm KS} <  10^{-3}$), which remains consistent with the findings of \citet{Heintz+20}. The majority of FRB hosts occupy the star-forming and LINER region of the BPT diagram. We note that all three hosts of repeating FRBs lie in the star-forming region and the host of repeating FRB20121102A remains an outlier as compared to the hosts of other FRBs \citep{Li+19}.
\begin{table*}
 \begin{threeparttable}
\centering
    \begin{tabular}{cc}
    \hline
    Transient type & Literature reference \\
   \hline
ULXs \tnote{a} & \citet{Kovlakas+20} \\
SGRBs & \citet{Leibler+10,Fong+10, Fong+13, Berger14} \\
LGRBs & \citet{Blanchard+16,Taggart+21} \\
CCSNe & \citet{Schulze+20} \\
SLSNe & \citet{Schulze+20,Taggart+21} \\
Type Ia SNe & \citet{Lampeitl+10,Uddin+20} \\

    \hline
    \end{tabular}
\caption{Literature references of the data used for various transients in this work. }
\label{tab:data}
\begin{tablenotes}
\item [a] Excluding unreliable data, nuclear sources and sources with X-ray luminosities $< 10^{39}$erg~s$^{-1}$.
\end{tablenotes}
\end{threeparttable}
\end{table*}

\section{Disentangling the host galaxies of repeating and non-repeating FRBs}
\label{sec:rep_oneoff}
The progenitors of FRBs are linked to the specific stellar population and environments of their host galaxies. Based on the larger set of FRBs and their hosts presented here, in combination with previous literature identifications, we leverage this larger sample to further constrain the likely progenitor channels of FRBs. More specifically, we aim to quantify whether the repeating and apparently non-repeating bursts are hosted by distinct galaxy environments. This might provide further clues to whether their progenitor channels are physically distinct. 

\begin{figure*}
\includegraphics[scale=0.44]{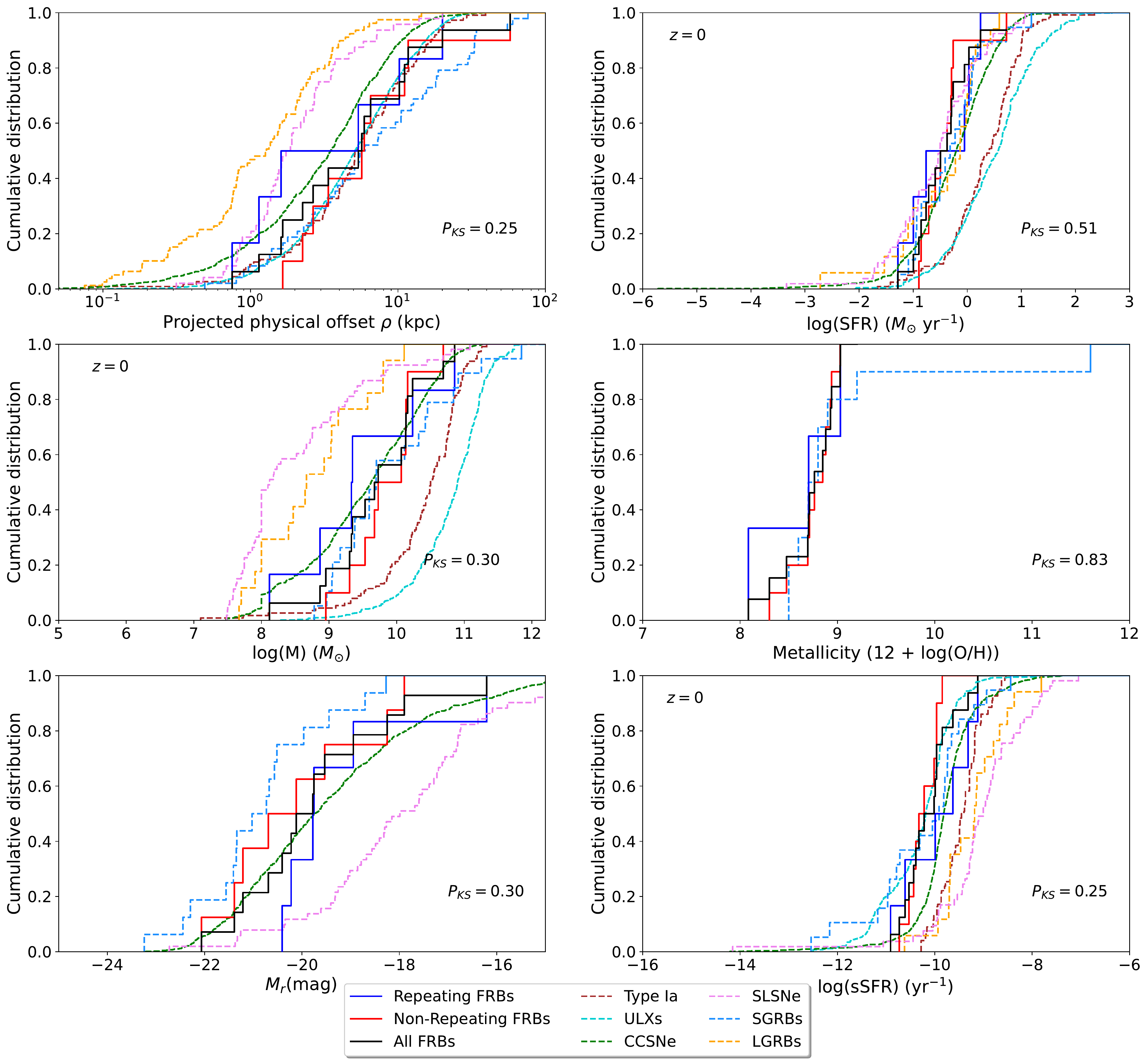}
\caption{Cumulative distributions of the projected physical offsets, SFR, stellar mass, metallicity, absolute $r$-band magnitudes and specific SFR for repeating (blue), non-repeating (red) and combined (black) FRB host population compared with ulta-luminous X-ray (ULX) sources (cyan), short (lightblue) and long (orange) gamma-ray bursts and core-collapse (CC)(green), superluminous (SL)(magenta) and Type Ia (brown) supernovae (SNe). The data for these transients are taken from literature listed in Table \ref{tab:data}. The p-values for KS-test between the repeating and non-repeating population is listed on the right side of each plot. We also summarize the p-values for KS-tests between FRB populations  and respective transients in Table \ref{tab:pvalues}. }
\label{fig:transients}
\end{figure*}

Initially, \citet{Heintz+20} found that FRB hosts seem to show an overall broad, continuous range of physical properties. They noted that the hosts of repeating FRBs generally occupied the faint, low-mass end of the FRB galaxy distribution. Here, we performed a differential analysis of the properties of the larger sample of FRB hosts. 
In Fig.\,\ref{fig:transients} we compared the projected physical offsets, SFR, stellar mass, metallicities, $r$-band luminosities and specific star formation rates of repeating and non-repeating FRB host population. We found the KS test p-values to be greater than our threshold statistical significance level ($\alpha = 0.05$). Thus, the null-hypothesis that these distributions are drawn from the same underlying distribution cannot be rejected.
The present sample of FRB hosts thus do not indicate strong physical distinctions between the two apparent source populations.

\definecolor{LightCyan}{rgb}{0.88,1,1}
\definecolor{Gray}{gray}{0.9}
\begin{deluxetable*}{cc|cccccc}
\tablewidth{0pc}
\tablecaption{P-values for two-sample 1D-KS after comparing the distributions of various properties (column 1) of the hosts of FRBs (all together and separated by repeating and on-off host population) with that of galaxies hosting ULXs, SGRBs, LGRBs, SLSNe, CCSNe and Type Ia SN. Highlighted cells shows the p-values ($>0.05$) for which the null hypothesis cannot be rejected. SGRBs and CCSNe remain as the
allowed dominant channels for
producing FRBs. \label{tab:pvalues}}
\tabletypesize{\footnotesize}
\tablehead{\colhead{Property} & \colhead{FRB Type} &\colhead{}  
& \colhead{} & \colhead{Transients} 
& \colhead{} 
& \colhead{}  
& \colhead{}
\\
\colhead{} & \colhead{} &\colhead{ULXs}  
& \colhead{SGRBs} & \colhead{LGRBs} 
& \colhead{SLSNe} 
& \colhead{CCSNe}  
& \colhead{Type Ia}
\\
} 
\startdata 
Offset & Rep & \cellcolor{Gray}0.28 &\cellcolor{Gray}0.44 & \cellcolor{Gray}0.18 & \cellcolor{Gray}0.37 & \cellcolor{Gray}0.82 & \cellcolor{Gray}0.28\\
& Non-rep & \cellcolor{Gray}0.97 & \cellcolor{Gray}0.57 & 0.00 & 0.02 & \cellcolor{Gray}0.25 & \cellcolor{Gray}0.99 \\
 & All & \cellcolor{Gray}0.93 & \cellcolor{Gray}0.52 & 0.00 & 0.02 & \cellcolor{Gray}0.27 & \cellcolor{Gray}0.92\\
\hline
SFR & Rep & 0.01 & \cellcolor{Gray}0.93 & \cellcolor{Gray}0.96 & \cellcolor{Gray}0.86 & \cellcolor{Gray}0.65 & 0.03 \\
& Non-rep & 0.00 & \cellcolor{Gray}0.26 & \cellcolor{Gray}0.07 & \cellcolor{Gray}0.18 & 0.04 & 0.00\\
 & All & 0.00 & \cellcolor{Gray}0.68 & \cellcolor{Gray}0.23 & \cellcolor{Gray}0.44 & \cellcolor{Gray}0.15 & 0.00\\
\hline
log(M) & Rep & 0.00 & \cellcolor{Gray}0.36 & \cellcolor{Gray}0.29 & \cellcolor{Gray}0.06 & \cellcolor{Gray}0.73 & 0.02 \\
& Non-rep & 0.00 & \cellcolor{Gray}0.64 & 0.00 & 0.00 & \cellcolor{Gray}0.33 & 0.00 \\
& All & 0.00 & \cellcolor{Gray}0.59 & 0.00 & 0.00 & \cellcolor{Gray}0.50 & 0.00  \\
\hline
Z & Rep & - & \cellcolor{Gray}0.91 & - & - & - & -  \\
& Non-rep & - & \cellcolor{Gray}0.99 & - & - & - & -  \\
 & All & - & \cellcolor{Gray}0.84 & - & - & -& - \\
\hline
M$_{\rm r}$ & Rep & - & 0.01 & - & \cellcolor{Gray}0.06 & \cellcolor{Gray}0.36& -  \\
& Non-rep & - & \cellcolor{Gray}0.89 & - & 0.01 & \cellcolor{Gray}0.80& -  \\
 & All & - & \cellcolor{Gray}0.06 & - & 0.00 & \cellcolor{Gray}0.73& - \\
\hline
sSFR & Rep & \cellcolor{Gray}0.32 & \cellcolor{Gray}0.75 & \cellcolor{Gray}0.25 & 0.05 & \cellcolor{Gray}0.70 & \cellcolor{Gray}0.34  \\
& Non-rep & \cellcolor{Gray}0.46 & \cellcolor{Gray}0.08 &  0.00 & 0.00 & 0.00& 0.00  \\
& All & \cellcolor{Gray}0.42 & \cellcolor{Gray}0.39 & 0.00 & 0.00 & 0.01& 0.00 \\
\hline 
\enddata 
\end{deluxetable*} 
\begin{figure*}
\begin{tabular}{cc}
\includegraphics[scale=0.48,trim={2cm 0.0cm 0.0cm 0.5cm}]{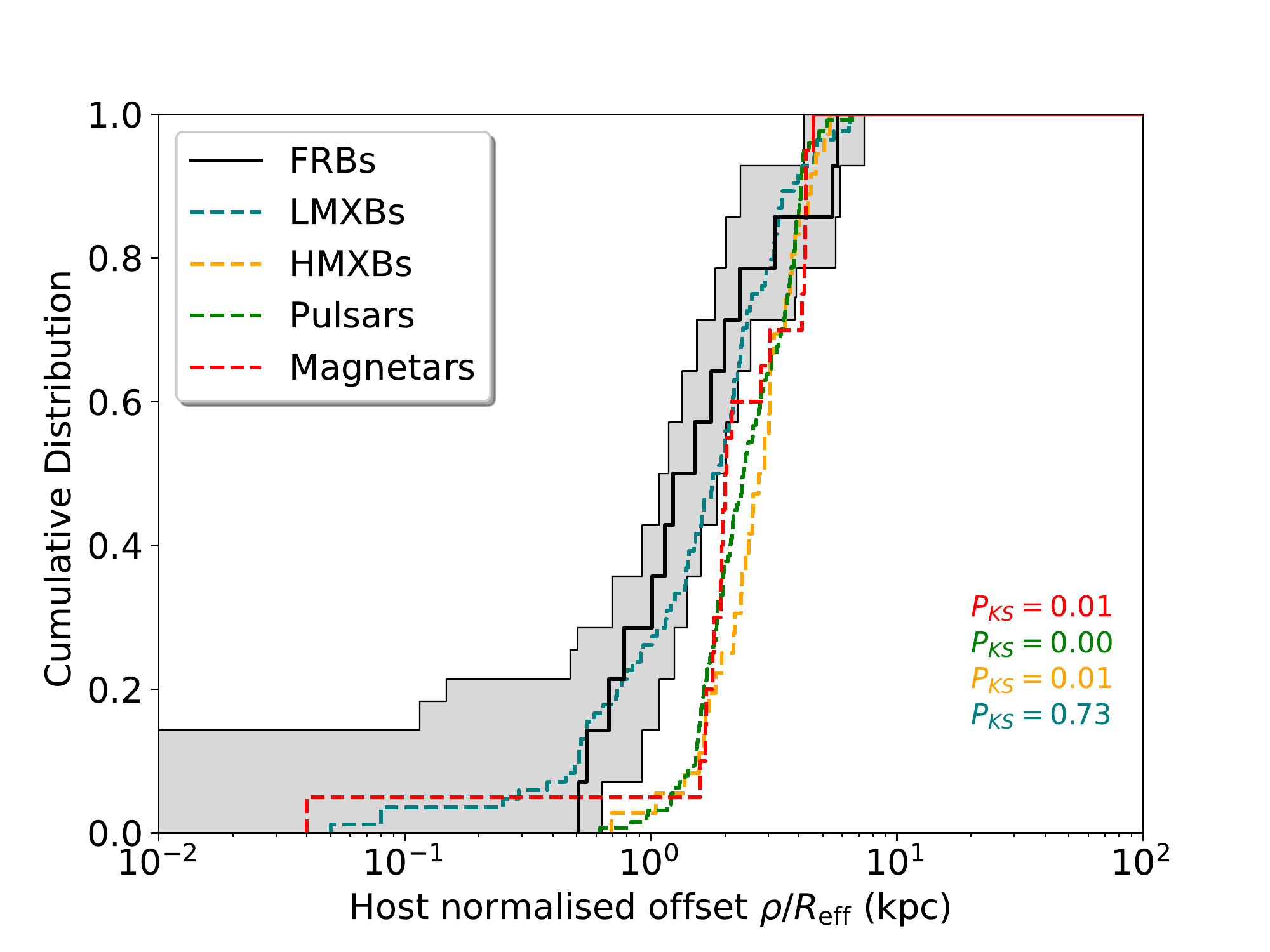} &
\includegraphics[scale=0.48,trim={2cm 0.0cm 0.0cm 0.5cm}]{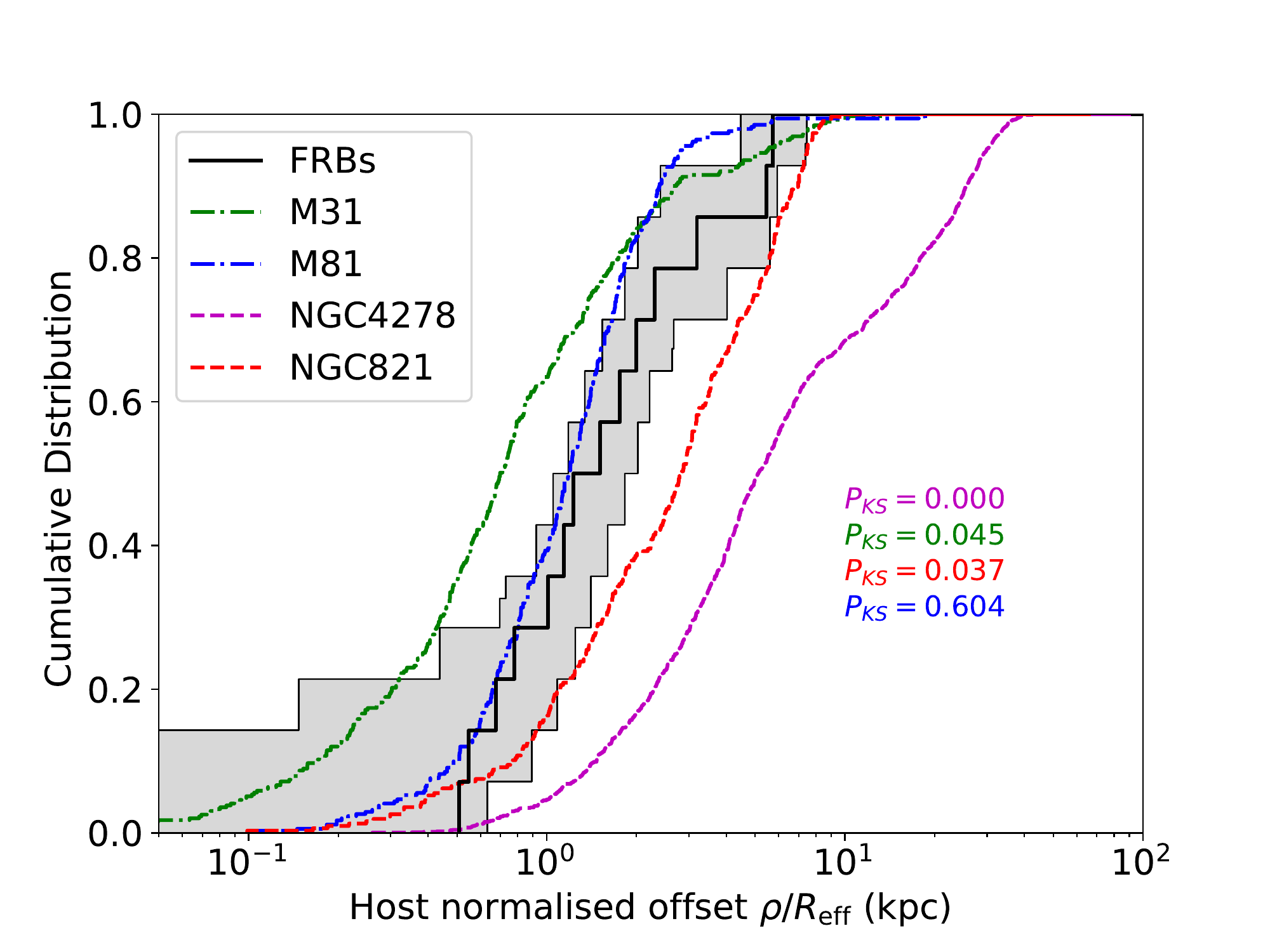}
\end{tabular}

\caption{Left: Cumulative distributions of the host-normalized offsets for FRBs compared with that of Galactic objects in the Milky Way galaxy. The data for various Galactic sources has been taken from \citet{Chrimes+21}. The gray-shaded region represents the $1\sigma$ uncertainty on the CDF, accounting for uncertainties due to individual measurements and sample size. The p-values for a KS-test between Galactic sources and all FRBs are listed on the right side of the plot. Right: Cumulative distributions of the host-normalized offsets for FRBs compared with offset distributions of globular clusters associated with M81, M31 (spiral late-type galaxies), NGC821 (isolated elliptical galaxy) and NGC4278 (L* elliptical galaxy). The effective radius used for these galaxies are $R_{\rm eff(M81)} = 3.5$~kpc \citep{Bhardwaj+21}, $R_{\rm eff(M31)} = 7.2$~kpc \citep{Savorgnan+16}, $R_{\rm eff(NGC821)} = 5.1$~kpc \citep{Pallegrini+07} and $R_{\rm eff(NGC4278)} = 2.4$~kpc \citep{Usher+13}. The p-values for KS-test between globular clusters and all FRBs are listed on the right side of the plot.}
\label{fig:MW}
\end{figure*}

\section{Comparison of the FRB host galaxy properties with other transients}
\label{sec:galactic}
\subsection{Extragalactic transients}

Studies of extragalactic transients often use their physical offsets from the center of their hosts, and locations within their host galaxies, to aid in inferring the nature of their progenitors. In the past, such investigations including other global properties for a small sample of FRB hosts have been conducted \citep{Bhandari_2020a,Heintz+20,Mannings+20,Ye+20,Safarzadeh+20,2021ApJ...907L..31B}. The majority of them suggest that galaxies hosting gamma-ray bursts (SGRBs) and CCSNe are similar to galaxies hosting FRBs. Additionally, the progenitor scenarios of LGRBs and SLSNe have been disfavored for the majority of FRBs. Recently, ultra-luminous X-ray sources (ULX)-like binaries were proposed as a possible progenitor for periodically active FRBs \citep{Sridhar+21}. Motivated by such studies, we performed two sample 1D-KS tests comparing the distributions of various global host properties of the most updated sample of FRBs together and separated into two populations (repeating and non-repeating) with the host galaxy properties of other transients (see Fig.\,\ref{fig:transients}). We also included Type Ia supernovae -- a proxy for the accretion-induced collapse of white dwarf stars, which have been suggested as possible FRB progenitors \citep{Margalit2019,Kirsten+21}. Table\,\ref{tab:data} represents the data from the literature that has been used in this study for all these transients. We used the method described in \citet{Bochenek2020} to correct for redshift evolution by scaling the stellar masses and star-formation rates of host galaxies of FRBs and other transients to be statistically representative of $z = 0$ galaxies.

While the distribution of projected physical offsets for ULXs and Type Ia supernovae are statistically consistent with that of FRBs, their host galaxies are often more massive and star-forming, i.e, a mean of log(M$_*$)~= 10.4~M$_{\odot}$ and 10.8~M$_{\odot}$, SFR = 6.2~M$_{\odot}$yr$^{-1}$ and 11.7~M$_{\odot}$yr$^{-1}$ for the hosts of Type Ia SN and ULXs, respectively. A possible caveat is that the ULX sample in the HECATE and the Chandra X-ray catalogue is biased against low-mass galaxies due to various selection effects \citep{Sridhar+21}. 
However, we observe the specific SFR for ULX hosts to be similar to that of FRB hosts.
Also, the majority of properties of the hosts of SGRBs and CCSNe are very similar to FRB hosts suggesting that the host population of FRB progenitors shares the same/similar characteristics to those of these transients.
A KS test shows that the SFR distribution of LGRB hosts and SLSNe hosts are consistent with FRB hosts. However, we note that such events are common in dwarf galaxies with high specific SFRs as evident from the left middle and right bottom panel of Fig.\,\ref{fig:transients}, which is not consistent with the overall FRB-host population. Interestingly, while the results of our KS-tests allow us to statistically rule out LGRB scenarios for all FRBs combined ($P_{\rm KS} < 0.05$), we are unable to reject the hypothesis that the repeating FRB and LGRB hosts are drawn from the same continuous distribution. When comparing with SLSNe hosts, we found the physical offsets of the repeating FRBs ($P_{\rm KSR} =0.37$), stellar mass ($P_{\rm KSR} =0.06$) and $r$-band magnitude ($P_{\rm KSR} =0.06$) of their hosts to agree with that of SLSNe hosts. The p-values from our KS-tests are presented in Table\,\ref{tab:pvalues}.

\subsection{Comparison to the Galactic source population and globular clusters}

The prevailing view currently associates FRBs with magnetars, whose extreme magnetic fields provide a reservoir of energy to produce FRBs \citep{Metzger2018, Lyutikov2019}. With the detection of an FRB-like burst from SGR1935+2154, the association of some (low-luminosity) FRBs with magnetars has been observationally confirmed \citep{Anderson2020,Bochenek2020}. Following the analysis of \citet{Chrimes+21}, we compared the host-normalized offset distributions of low-mass X-ray binaries (LMXBs), high-mass X-ray binaries (HMXBs), pulsars and magnetars in the Milky Way with the updated FRB host sample --- including all repeating and non-repeating FRBs in the left panel of Fig.\,\ref{fig:MW}. Based on our KS-test, the offset distribution of only low-mass X-ray binaries in the Milky Way is consistent with that observed for FRBs in their host galaxies (95\% confidence).

Recently, a repeating FRB20200120E originally associated with the galaxy M81 \citep{Bhardwaj+21} has been precisely localized, by the EVN network, to a globular cluster (GC) system [PR95] 30244 in M81 \citep{Kirsten+21}. Motivated by this finding, we compared the host-normalized offset distributions of FRBs in our sample with globular clusters in late-type spirals and early-type elliptical galaxies in the right panel of Fig.\,\ref{fig:MW}. 

For the GCs in late-type galaxies, we used a sample of 340 high quality GC candidates (of which 74 were confirmed using spectroscopy) associated with M81, identified in HST imaging \citep{Nantais+10a,Nantais+10b}. We note that
the available data is dominated by a GC disk population with the majority of GCs within 10~kpc from the center of M81. We also used a sample of 390 GCs associated with the galaxy M31 (both disk and halo population) identified in images from the Wide Field Camera (WFCAM) on the United Kingdom Infrared Telescope and from the Sloan Digital Sky Survey (SDSS) \citep{Peacock+10}. For early-type galaxies, we used the available GC data associated with an isolated elliptical galaxy NGC821, extending up to 50~kpc from the center of the galaxy \citep{Spitler+08}. The sample consist of 306 GCs identified in the new imaging from the 3.5-m Wisconsin Indiana Yale NOAO (WIYN) Mini-Mosaic imager, supplemented with the HST WFPC2 images.
We also used a sample of 1828 GC candidates (of which 270 are confirmed using spectroscopy from Keck/DEIMOS) associated with the $L_{*}$ elliptical galaxy NGC4278, identified in HST/ACS and wide-field Subaru/Suprime-Cam imaging \citep{Usher+13}. These selected galaxies have stellar masses and SFRs in the range  M$_\star = 10^{10.5} - 10^{11.5}\,$M$_{\odot}$ and SFR = $0.01-1\,$M$_{\odot}$\,yr$^{-1}$, overlapping with the SFRs and high-mass end distribution of FRB host galaxies \citep{M31_SM,M31_SFR,Bhardwaj+21,Forbes+16,AKARI}.

We compared the host-normalized offset distribution of FRBs in their hosts with that of globular clusters in selected galaxies and found that they are not consistent with being drawn from the same underlying distribution (95\% confidence) except for M81 ($P_{\rm KS} = 0.604$). This could be due to the fact that M81 globular cluster data is incomplete and dominated by disc population.

A possible caveat in above analysis is that the number of GCs scale with galaxy stellar mass and our FRB host population spans four orders of magnitude in the stellar mass, i.e.\ M$_\star = 10^{8} - 10^{11}\,$M$_{\odot}$. We also repeated the KS-test by selecting FRB hosts which had M$_\star>10^{10}\,$M$_{\odot}$ and found similar results, however with lower confidence of 80\%. Furthermore, we note that our FRB host sample currently shows a deficiency of elliptical galaxies.
A subset of FRBs in our sample were found to originate from or near the spiral arms of their hosts. Whether they are linked to the disk population of GCs is presently unknown and we are limited by the sensitivity of current telescopes. A much larger sample of precisely localized FRBs with high spatial observations of nearby hosts are needed to pursue this further. 

\section{Summary and future work}
\label{sec:summary}
We have presented the localization of the sixth repeating FRB20180301A using the \rf\ system at the VLA and two apparently non-repeating FRBs (FRB20191228A and FRB20200906A) discovered by ASKAP. With an updated sample of 6 repeating and 10 non-repeating FRB host galaxies, we have conducted a differential analysis of global properties of the FRB host population. While the latest observations of FRBs from the CHIME/FRB project strongly suggest that repeaters and single-burst sources arise from separate mechanisms and astrophysical sources \citep{Pleunis+21}, we did not find significant differences in their host populations. We observed FRB hosts to be moderately star-forming galaxies ($0.06-8$\,M$_{\odot}\,{\rm yr}^{-1}$), with masses offset from the star-forming main-sequence. The majority of FRB hosts lie in the star-forming and LINER region of the BPT diagram. 
As a low metallicity dwarf galaxy, the host of FRB20121102A continues to be an outlier in the sample.  Furthermore, we observe no persistent radio emission co-located with the bursts in our radio follow-up observations of FRBs\,20180301A, 20191228A, and 20200906A. We note that the derived $3\sigma$ upper limits on the luminosity of these sources are lower than the luminosity of the FRB20121102A persistent source ($2.1 \times 10^{22}$\,W/Hz at 1.4\,GHz \citep{Ofek2017}) indicating that these bursts might originate from less extreme environments.

We find that FRBs in our sample do not track the stellar mass and in general are not hosted in old, red and dead galaxies which have old stellar population. The dearth of FRBs in the massive red galaxies suggest that FRBs are not solely produced in channels with a large average delay between star formation and the FRB source formation such as magnetars formed via compact object-related systems, including neutron star mergers, or the AIC of a white dwarf to a neutron star. Current data supports a mix of prompt (core-collapse SNe) and delayed channels for producing FRB progenitors, suggesting that they are drawn from the general stellar population rather than an exotic and rare sub-population. Furthermore, FRB hosts do not follow the specific SFRs of main-sequence star-forming galaxies, nor do they track the star formation rates of field galaxies in the nearby Universe.

When comparing the properties of FRB host galaxies with that of other transients, we find the host galaxies of ULXs and Type Ia supernovae to be more massive and star-forming than FRBs. The hosts of CCSNe and SGRBs are similar to FRB hosts in terms of their stellar masses, star-formation rates, projected physical offsets, absolute $r$-band magnitudes, and specific SFRs. While from the host galaxy considerations, we could statistically rule out LGRBs and SLSNe as progenitor scenarios for all FRBs combined, we found some similarities between the repeating FRB host population and hosts of LGRBs and SLSNe. These may be attributed to either the small sample size or the effect of the outlier FRB20121102A host galaxy on the overall repeating FRB host population. We also note that one is more likely to find an extreme value under the null hypothesis with more KS-tests.

Driven by the studies and findings of \citet{Chrimes+21} and \citet{Kirsten+21}, we compared the physical offsets of FRBs in their hosts with that of Galactic sources such as pulsars, magnetars, X-ray binaries in the Milky Way, and globular clusters in the late- and early-type galaxies. According to our KS test, the Galactic source offset distributions of the neutron star population and high mass X-ray binaries are different, while low-mass X-ray binaries are indistinguishable from the observed FRB offset distribution (95\% confidence). We also show that FRBs are positioned in their host galaxies in a way that is mostly not comparable to globular clusters found in late-type spiral and early-type elliptical galaxies.

Lastly, in the future, observations of a much larger sample of nearby FRB hosts will be ideal for progenitor model studies as these will allow the high spatial resolution analysis of FRB environments in their host galaxies. 

All of the data and the majority of the software
used for the host analysis is available at \linebreak
https://github.com/FRBs/FRB.

\software{PYSE \citep{Pyse}, ESOReflex \citep{ESOReflex}, Montage \citep{Montage}, Source-Extractor \citep{SExtractor}, Astropy \citep{astropy1}, Burstfit \citep{2021arXiv210705658A}, Scipy \citep{scipy}, CASA \citep{CASA}, CIGALE \citep{cigale}, pPXF \citep{ppxf}}.
\\
\\
\noindent 
SB would like to thank Themiya Nanayakkara, Arash Bahramian, and Kristen Dage for useful discussions. 
RMS acknowledges support the Australian Research Council Future Fellowship FT190100155. KEH acknowledges support by a Postdoctoral Fellowship Grant (217690--051) from The Icelandic Research Fund. K.A. acknowledges support from NSF grant AAG-1714897. L.M. acknowledges the receipt of an MQ-RES scholarship from Macquarie University. S.B.S acknowledges support from NSF grant AAG-1714897. She is a CIFAR Azrieli Global Scholar in the Gravity and the Extreme Universe program, which helped support J.S. for this project. ATD is the recipient of an Australian Research Council Future Fellowship (FT150100415). W.F. acknowledges support by the National Science Foundation under grant Nos. AST-1814782, AST-1909358 and CAREER grant No. AST-2047919. CJL acknowledges support from the National Science Foundation under Grant No.\ 2022546. NT acknowledges support by FONDECYT grant 11191217. KJL is supported by CAS XDB23010200, Max-Planck Partner Group, National SKA program of China 2020SKA0120100, NSFC 11690024, CAS Cultivation Project for FAST Scientific. The NANOGrav project receives support from National Science Foundation (NSF) Physics Frontiers Center award number 1430284. Part of this research was carried out at the Jet Propulsion Laboratory, California Institute of Technology, under a contract with the National Aeronautics and Space Administration. Authors S.S., N.T., J.X.P., and K.G.L. as members of the Fast and Fortunate for FRB
Follow-up team, acknowledge support from 
NSF grants AST-1911140 and AST-1910471.
The Australian Square Kilometre Array Pathfinder and Australia Telescope Compact Array (ATCA) are part of the Australia Telescope National Facility which is managed by CSIRO. 
Operation of ASKAP is funded by the Australian Government with support from the National Collaborative Research Infrastructure Strategy. ASKAP uses the resources of the Pawsey Supercomputing Centre. Establishment of ASKAP, the Murchison Radio-astronomy Observatory and the Pawsey Supercomputing Centre are initiatives of the Australian Government, with support from the Government of Western Australia and the Science and Industry Endowment Fund. 
We acknowledge the Wajarri Yamatji as the traditional owners of the Murchison Radio-astronomy Observatory site. 
We acknowledge the Gomeroi people as the traditional owners of the Paul Wild (ATCA) Observatory site.
The National Radio Astronomy Observatory is a facility of the National Science Foundation operated under cooperative agreement by Associated Universities, Inc. 
Spectra were obtained at the W. M. Keck Observatory, which is operated as a scientific partnership among Caltech, the University of California, and the National Aeronautics and Space Administration (NASA). W. M. Keck Observatory and MMT Observatory access was in part supported by Northwestern University and the Centre for Interdisciplinary Exploration and Research in Astrophysics (CIERA). The Keck Observatory was made possible by the generous financial support of the W. M. Keck Foundation. The authors recognize and acknowledge the very significant cultural role and reverence that the summit of Mauna Kea has always had within the indigenous Hawaiian community. We are most fortunate to have the opportunity to conduct observations from this mountain. Observations reported here were obtained at the MMT Observatory, a joint facility of the University of Arizona and the Smithsonian Institution.
Based on observations collected at the European Southern Observatory under ESO programme 0105.A-0687(A).
Based on observations obtained at the international Gemini Observatory, a program of NSF’s NOIRLab, which is managed by the Association of Universities for Research in Astronomy (AURA) under a cooperative agreement with the National Science Foundation on behalf of the Gemini Observatory partnership: the National Science Foundation (United States), National Research Council (Canada), Agencia Nacional de Investigaci\'{o}n y Desarrollo (Chile), Ministerio de Ciencia, Tecnolog\'{i}a e Innovaci\'{o}n (Argentina), Minist\'{e}rio da Ci\^{e}ncia, Tecnologia, Inova\c{c}\~{o}es e Comunica\c{c}\~{o}es (Brazil), and Korea Astronomy and Space Science Institute (Republic of Korea). The Gemini data were obtained from program GS-2020B-Q-138, and were processed using
the DRAGONS (Data Reduction for Astronomy from Gemini Observatory North and South) package.

\bibliography{references.bib}
\bibliographystyle{aasjournal}

\appendix 
\setcounter{table}{0}
\renewcommand{\thetable}{A\arabic{table}}
\renewcommand{\thefigure}{A\arabic{figure}}    
\setcounter{figure}{0}    
\noindent
The spectrum of the new FRB hosts in this paper and newly obtained data for FRB20200430A are shown in Fig.\,\ref{fig:spectrum}. Table\,\ref{tab:spectrum} lists the nebular line measurements of FRB hosts that are used in the BPT diagram.  We have also updated the previously reported Pan-STARRS photometric measurements (PSFMag) for hosts of FRB20190523A, FRB20190714A, and FRB20200430A to Kron Mag and re-did the SED fitting in \texttt{CIGALE} using the method described in \citet{Heintz+20}. Furthermore, we obtained new VLT/FORS2 measurements for the host of FRB20190611B. The best fitting models for these hosts and new FRBs introduced in this work are presented in Fig.\,\ref{fig:cigale} and the updated derived measurements are shown in Table\,\ref{tab:updated_properties}. The new and updated photometry is presented in Table\,\ref{tab:photometry}.
\begin{figure*}
\begin{tabular}{cc}
\includegraphics[scale=0.62,trim={00cm 0cm 0cm 0.5cm}]{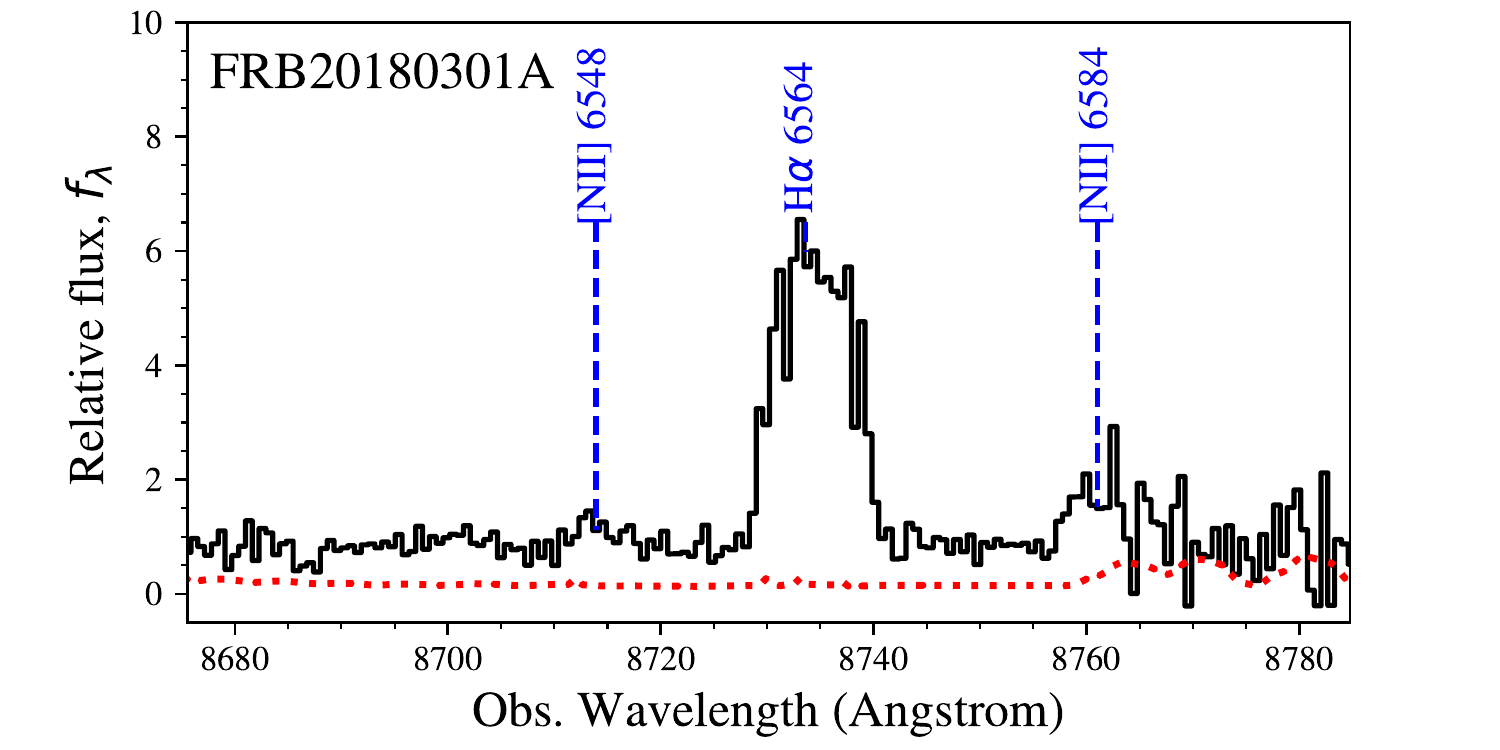}
\includegraphics[scale=0.62,trim={0cm 0cm 0cm 0.5cm}]{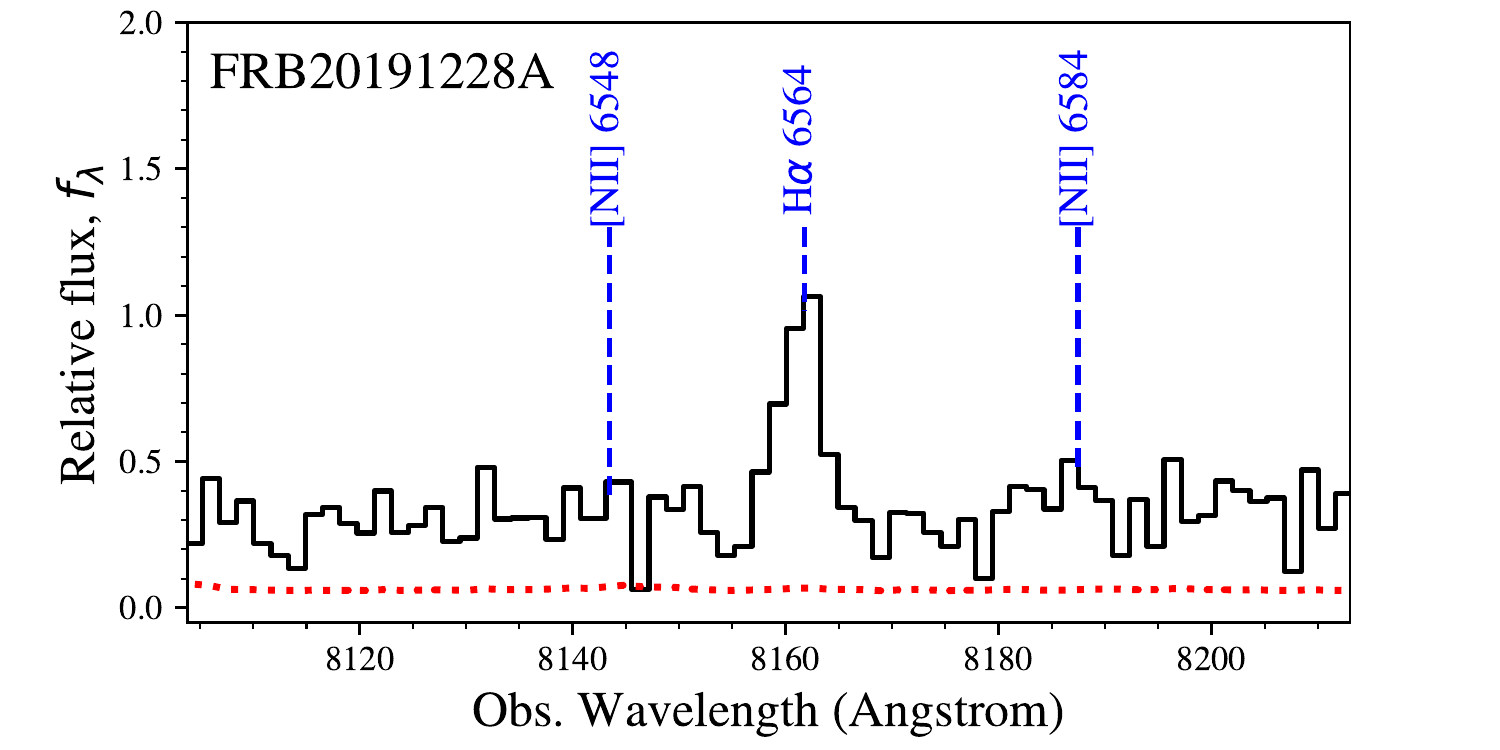} \\
\includegraphics[scale=0.62,trim={0cm 0cm 0cm 0cm}]{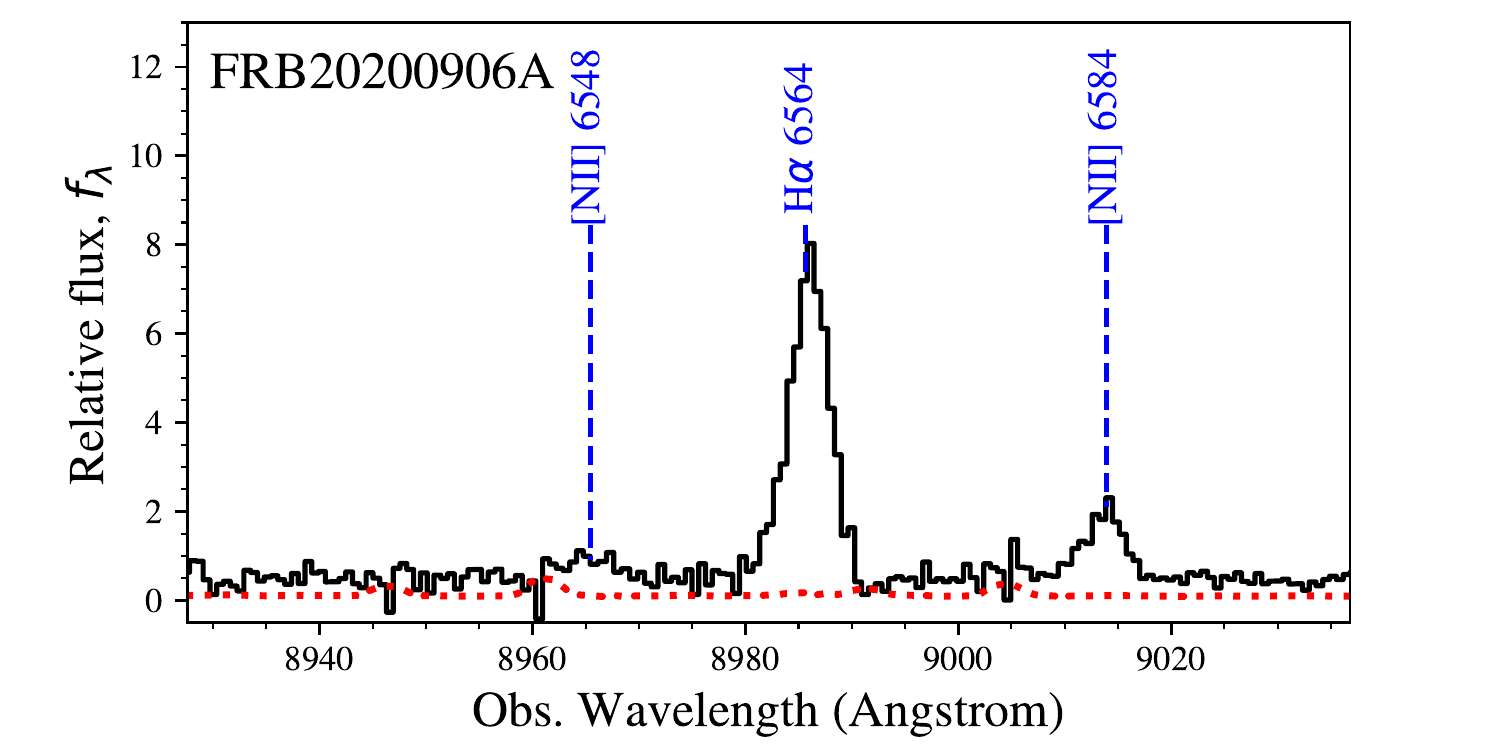}
\includegraphics[scale=0.62,trim={0cm 0cm 0cm 0cm}]{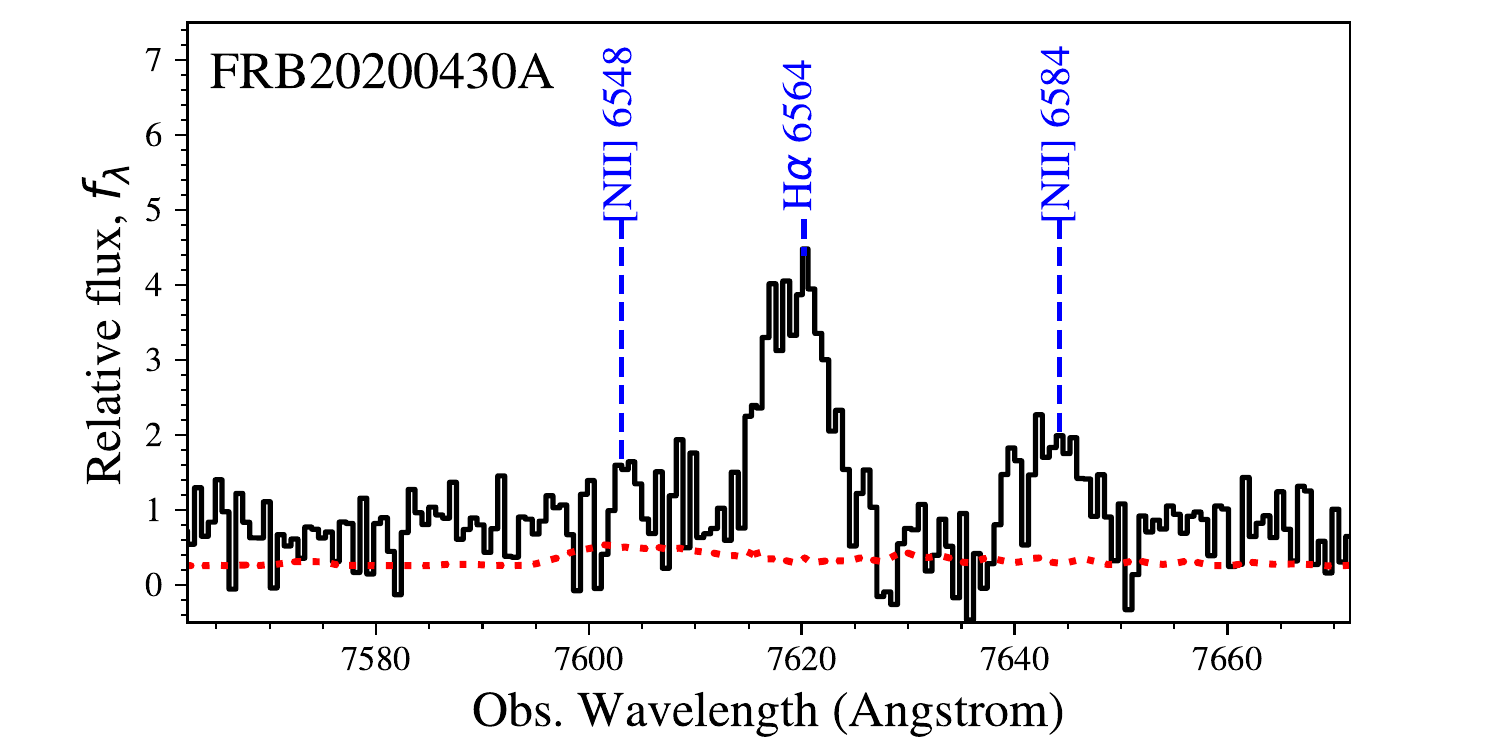}
\end{tabular}
\caption{A zoomed in version of the optical spectrum of the host of FRB20180301A, FRB20191228A, FRB20200906A and FRB20200430A.}
\label{fig:spectrum}
\end{figure*}

\definecolor{LightCyan}{rgb}{0.88,1,1}
\definecolor{Gray}{gray}{0.9}
\begin{deluxetable*}{ccccc}
\tablewidth{0pc}
\tablecaption{Nebular line emission measurements in units of 10$^{-16}$~erg~s$^{-1}$~cm$^{-2}$. New FRB hosts presented in this paper are highlighted in cyan and old FRB hosts with new data are highlighted in gray. The data for FRB20171020A and FRB20201124A is taken from \citet{Mahony2018} and \citet{Fong+21} respectively. \label{tab:spectrum}}
\tabletypesize{\footnotesize}
\tablehead{\colhead{FRB}  
& \colhead{$H\alpha$} & \colhead{$H\beta$} 
& \colhead{[N~{\sc ii}]} 
& \colhead{[O~{\sc iii}]}  
\\} 
\startdata 
FRB20121102A  &$2.61 \pm0.04$ &$0.96 \pm0.09$ &$<0.12$ &$4.38 \pm0.08$ \\
\rowcolor{LightCyan}
FRB20180301A  &$16.97 \pm0.18$ &$7.77 \pm0.61$ &$4.32 \pm0.32$ &$12.75 \pm0.65$ \\
FRB20180916B  &$40.27 \pm0.25$ &- &$15.24 \pm0.24$ &$71.62 \pm0.60$ \\
FRB20180924B  &$2.79 \pm0.03$ &$0.72 \pm0.02$ &$1.94 \pm0.03$ &$0.79 \pm0.02$ \\
FRB20190102C  &$5.66 \pm0.17$ &$1.90 \pm0.17$ &$1.69 \pm0.19$ &$3.80 \pm0.27$ \\
FRB20190608B  &$27.65 \pm0.41$ &$8.37 \pm0.33$ &$18.32 \pm0.38$ &$14.95 \pm0.44$ \\
FRB20190611B  &$0.49 \pm0.05$ &$0.12 \pm0.03$ &$0.12 \pm0.04$ &$0.18 \pm0.04$ \\
FRB20190711A  &- &$0.26 \pm0.05$ &- &- \\
FRB20190714A  &$3.89 \pm0.03$ &$0.97 \pm0.03$ &$1.70 \pm0.03$ &$0.31 \pm0.03$ \\
FRB20191001A  &$27.38 \pm0.26$ &$5.01 \pm0.30$ &$13.91 \pm0.19$ &$3.62 \pm0.35$ \\
\rowcolor{LightCyan}
FRB20191228A  &$0.30 \pm0.02$ &$0.02 \pm0.02$ &$0.08 \pm0.02$ &$0.13 \pm0.03$ \\
\rowcolor{Gray}
FRB20200430A  &$4.27 \pm0.15$ &$1.15 \pm0.21$ &$2.02 \pm0.20$ &$1.17 \pm0.25$ \\
\rowcolor{LightCyan}
FRB20200906A  &$6.49 \pm0.07$ &$4.25 \pm0.14$ &$1.80 \pm0.06$ &$5.38 \pm0.17$ \\
FRB20201124A & 56.9$^{+14.9}_{-9.9}$ & 12.3$^{+3.0}_{-2.3}$ & 25.3$^{+6.0}_{-4.7}$ & 3.8$^{+2.1}_{-1.1}$ \\
\hline 
\enddata 
\end{deluxetable*} 

\begin{deluxetable*}{cccccccccccccccc}
\tablewidth{0pc}
\tablecaption{Results for FRB Associations\label{tab:results}}
\tabletypesize{\footnotesize}
\tablehead{\colhead{FRB}  
& \colhead{RA$_{\rm cand}$} & \colhead{Dec$_{\rm cand}$} 
& \colhead{$\theta$} 
& \colhead{\halflight}  
& \colhead{\gmag} 
& \colhead{Filter} & \colhead{\pchance} 
& \colhead{\PO} & \colhead{\POx} 
\\} 
\startdata 
FRB20121102A& 82.9945 & $33.1479$& 0.2& 0.28& 23.52& GMOS\_N\_i& 0.0039& 0.0245& 1.0000\\ 
FRB20180916A& 29.5012 & $65.7148$& 7.7& 3.03& 16.16& GMOS\_N\_r& 0.0005& 0.8200& 1.0000\\ 
FRB20180924B& 326.1054 & $-40.9002$& 0.8& 1.31& 21.32& VLT\_FORS2\_g& 0.0119& 0.8723& 0.9894\\ 
FRB20181112A& 327.3486 & $-52.9709$& 0.4& 0.67& 21.49& VLT\_FORS2\_I& 0.0622& 0.0784& 0.6678\\ 
FRB20190102C& 322.4149 & $-79.4756$& 0.5& 0.86& 20.73& VLT\_FORS2\_I& 0.0056& 0.8425& 1.0000\\ 
FRB20190523A& 207.0642 & $72.4706$& 3.4& 0.71& 22.13& LRIS\_R& 0.1158& 0.1974& 0.8154\\ 
FRB20190608B& 334.0203 & $-7.8988$& 2.5& 1.66& 17.60& VLT\_FORS2\_I& 0.0005& 0.9930& 1.0000\\ 
FRB20190611B& 320.7429 & $-79.3973$& 2.0& 0.50& 22.35& GMOS\_S\_i& 0.0407& 0.3322& 0.9741\\ 
FRB20190614D& 65.0743 & $73.7068$& 1.3& 0.41& 24.01& LRIS\_I& 0.0552& 0.1944& 0.5825\\ 
FRB20190711A& 329.4194 & $-80.3581$& 0.5& 0.46& 22.93& GMOS\_S\_i& 0.0172& 0.4782& 0.9937\\ 
FRB20190714A& 183.9795 & $-13.0212$& 1.2& 0.95& 19.48& VLT\_FORS2\_I& 0.0014& 0.7993& 1.0000\\ 
FRB20191001A& 323.3525 & $-54.7487$& 4.2& 1.36& 17.82& VLT\_FORS2\_I& 0.0010& 0.5075& 0.5980\\ 
FRB20200430A& 229.7064 & $12.3766$& 0.4& 0.72& 21.18& LRIS\_I& 0.0047& 0.9379& 1.0000\\ 
FRB20191228A& 344.4307 & $-29.5940$& 2.1& 0.49& 21.92& VLT\_FORS2\_I& 0.0251& 0.5596& 1.0000\\ 
FRB20200906A& 53.4958 & $-14.0833$& 1.6& 1.51& 20.70& VLT\_FORS2\_g& 0.0106& 0.8997& 1.0000\\ 
FRB20180301A& 93.2269 & $4.6704$& 2.3& 0.92& 22.07& GMOS\_S\_r& 0.0387& 0.7154& 0.9993\\ 
\hline 
\enddata 
\end{deluxetable*} 

\definecolor{LightCyan}{rgb}{0.88,1,1}
\definecolor{Gray}{gray}{0.9}

\begin{deluxetable*}{cccccccccccc}

\tablewidth{0pc}
\tablecaption{Host galaxies of six repeating and ten non-repeating FRBs used in this work, for which PATH posterior probability is $>90\%$. Derived measurements for new FRB hosts are highlighted in cyan. Updated measurement of previously reported hosts in \citet{Heintz+20} are highlighted in gray. We have also added most probable host of FRB20171020A \citep{Mahony2018} in our sample including the published host of repeating FRB20200120E \citep{Bhardwaj+21} and FRB20201124A \citep{Fong+21}.
\label{tab:updated_properties}}
\tabletypesize{\footnotesize}

\tablehead{\colhead{S.No} &\colhead{FRB}  
& \colhead{z} & \colhead{Rep} 
& \colhead{Offset} 
& \colhead{R$_{\rm eff}$\tablenotemark{a}}
& \colhead{Mass}  
& \colhead{SFR} 
& \colhead{log(sSFR)} 
& \colhead{$M_{\rm r}$} &  \colhead{$u-r$} 
& \colhead{$Z$}  
\\
\colhead{} & \colhead{}
& \colhead{} & \colhead{} 
& \colhead{(kpc)} 
& \colhead{(kpc)}
& \colhead{(10$^{10}~M_{\odot}$)}  
& \colhead{($M_{\odot}$ yr$^{-1}$)} 
& \colhead{(yr$^{-1}$)} 
& \colhead{} &  \colhead{} 
&  \colhead{} 
\\
} 
\startdata 
1 &FRB20121102A & 0.1927 & y & $0.8 \pm0.1$ & $0.66\pm0.03$ & $0.01 \pm0.01$ & $0.15 \pm0.04$ & $-$8.99 & $-16.20 \pm0.08$ & $1.49 \pm0.18$  & 8.08  \\
\rowcolor{LightCyan}
2 & FRB20180301A & 0.3304 & y & $10.2 \pm3.0$ & $5.80 \pm 0.20$ & $0.23 \pm0.06$ & $1.93 \pm0.58$ & $-$9.08 & $-20.18 \pm0.07$ & $0.90 \pm0.11$  & 8.70  \\
3 & FRB20180916B & 0.0337 & y & $5.4 \pm0.0$ & $3.57\pm 0.36$& $0.22 \pm0.03$ & $0.06 \pm0.02$ & $-$10.58 & $-19.46 \pm0.05$ & $1.53 \pm0.06$  &  -  \\
4 & FRB20180924B & 0.3212 & n & $3.4 \pm0.8$ & $2.75\pm 0.10$ &$1.32 \pm0.51$ & $0.88 \pm0.26$ & $-$10.18 & $-20.81 \pm0.05$ & $1.78 \pm0.15$  & 8.93  \\
5 & FRB20190102C & 0.2912 & n & $2.3 \pm4.2$ & $4.43 \pm 0.51$ &$0.47 \pm0.54$ & $0.86 \pm0.26$ & $-$9.74 & $-19.87 \pm0.06$ & $1.44 \pm0.15$  & 8.70  \\
6 & FRB20190608B & 0.1178 & n & $6.5 \pm0.8$ & $2.84 \pm 0.23$ &$1.16 \pm0.28$ & $0.69 \pm0.21$ & $-$10.22 & $-21.22 \pm0.05$ & $1.40 \pm0.09$  & 8.85  \\
\rowcolor{Gray}
7 & FRB20190611B & 0.3778 & n & $11.7 \pm5.8$ & $2.15 \pm 0.11$ &$0.35 \pm0.70$ & $0.27 \pm0.08$ & $-$10.11 & $-19.29 \pm0.10$ & $1.29 \pm0.18$  & 8.71  \\
8 & FRB20190711A & 0.5220 & y & $1.6 \pm4.5$ & $2.94\pm 0.17$ &$ 0.08 \pm0.03$ & $0.42 \pm0.12$ & $-$9.29 & $-19.01 \pm0.08$ & $0.95 \pm0.16$  &  -  \\
\rowcolor{Gray}
9 & FRB20190714A & 0.2365 & n & $2.7 \pm1.8$ & $3.94\pm 0.05$ &$1.42 \pm0.55$ & $0.65 \pm0.20$ & $-$10.34 & $-20.37 \pm0.05$ & $1.51 \pm0.28$  & 9.03  \\
10 & FRB20191001A & 0.2340 & n & $11.1 \pm0.8$ & $5.55\pm 0.03$ & $4.64 \pm1.88$ & $8.06 \pm2.42$ & $-$9.76 & $-22.13 \pm0.05$ & $1.67 \pm0.19$  & 8.94  \\
\rowcolor{LightCyan}
11 & FRB20191228A & 0.2432 & n & $5.7 \pm3.3$ & $1.78 \pm 0.06$ & $0.54 \pm0.60$ & $0.50 \pm0.15$ & $-$10.03 & $-18.26 \pm0.05$ & $2.13 \pm0.74$  & 8.48 \\
\rowcolor{Gray}
12 & FRB20200430A & 0.1608 & n & $1.7 \pm2.2$ & $1.64 \pm 0.53$ & $0.21 \pm0.11$ & $0.26 \pm0.08$ & $-$9.89 & $-18.25 \pm0.05$ & $2.08 \pm0.30$  & 8.88  \\
\rowcolor{LightCyan}
13 & FRB20200906A & 0.3688 & n & $5.9 \pm2.0$ & $7.58 \pm 0.06$ & $1.33 \pm0.37$ & $0.48 \pm0.14$ & $-$10.44 & $-21.49 \pm0.05$ & $1.22 \pm0.11$  & 8.76  \\
14 & FRB20171020A\tablenotemark{*} & 0.0087 & n &  - & - & 0.09 & 0.13 & $-$9.84 & $-$17.9  & - & 8.30 \\
15 & FRB20200120E\tablenotemark{*} & 0.0008 & y & $20.1 \pm3.0$ & 3.5 & $7.20 \pm1.70$ & $0.89 \pm0.27$ & $-$10.91 & $-$19.78  & $2.77 \pm0.00$ & - \\
16 & FRB20201124A\tablenotemark{*} & 0.0980 & y & $1.3\pm0.1$ & - & $1.70^{+0.08}_{-0.11}$ & $2.12^{+0.69}_{-0.28}$ & $-$9.90 & $-20.41 \pm0.03$ & $2.06 \pm0.55$  & 9.03 \\
\rowcolor{Gray}
-- &FRB20190523A & 0.6600 & n & $27.2 \pm22.6$ & $3.28 \pm 0.18$ & $46.49 \pm35.51$ & $0.09 \pm0.03$ & $-$12.74 & $-22.69 \pm0.14$ & $2.20 \pm0.22$  &  -  \\
\hline 
\enddata 
\tablenotetext{a}{Isophotal effective half-light radius are used for a sample of FRB host published in \citet{Mannings+20}. For remaining hosts, effective radius derived from \texttt{GALFIT} are used. }
\tablenotetext{*}{Host properties are taken from \citet{Mahony2018, Bhardwaj+21, Fong+21}. The observed magnitudes for FRB20171020A, FRB20200120E and FRB20201124A are approximated as rest-frame magnitudes because of their low redshifts. }
\end{deluxetable*} 

\definecolor{LightCyan}{rgb}{0.88,1,1}
\definecolor{Gray}{gray}{0.9}

\begin{deluxetable*}{cccccccc}

\tablewidth{0pc}
\tablecaption{Photometric measurements for a sample of new FRBs in this work and updated Pan-STARRS measurements for old FRBs published in \citet{Heintz+20}. All photometry have been corrected for Galactic extinction. 
\label{tab:photometry}}
\tabletypesize{\footnotesize}
\tablehead{\colhead{Filter} &\colhead{FRB20180301A}  
& \colhead{FRB20191228A} & \colhead{FRB20200906A} 
& \colhead{FRB20190523A} 
& \colhead{FRB20190611B} 
& \colhead{FRB20190714A}
& \colhead{FRB20200430A} \\
} 
\startdata 
NOT$_u$ & $21.46\pm0.30$ & - & - & - & - & - & - \\
NOT$_g$ & $21.40\pm0.09$ & - & - & - & - & - & -\\
NOT$_r$ & $21.04\pm0.06$ & - & - &- & - & - & -\\
NOT$_i$ & $20.99\pm0.06$ & - & - &- & - & - & -\\
NOT$_z$ & $20.59\pm0.11$ & - & - &- & - & - & -\\
Pan-STARRS$_g$ & - &  - & $20.93\pm0.05$ & $22.10\pm0.14$ & - & $20.88\pm0.04$ & $22.21\pm0.10$\\
Pan-STARRS$_r$ & - &  - & $20.15\pm0.03$ & $21.91\pm0.15$ & - & $20.32\pm0.03$ & $21.25\pm0.07$\\
Pan-STARRS$_i$ & - &  - & $19.89\pm0.02$ & $20.86\pm0.08$ & - & $19.82\pm0.02$ & $20.97\pm0.05$\\
Pan-STARRS$_z$ & - &  - & $19.62\pm0.03$ & $20.63\pm0.09$ & - & $19.63\pm0.03$ & $20.71\pm0.10$\\
Pan-STARRS$_y$ & - &  - & $19.65\pm0.08$ & $19.88\pm0.11$ & - & $19.43\pm0.06$ & -\\
VLT/FORS2$_g$ & - & $22.61\pm0.50$  & $20.54\pm0.05$ & - & $23.24 \pm 0.09$ & \citet{Heintz+20} & - \\
VLT/FORS2$_I$ & - & $21.96\pm0.40$  & $19.51\pm0.02$ & - & $22.02 \pm 0.07$ & \citet{Heintz+20} & - \\
MMT/MMRS$_J$ & $20.56\pm0.07$ & - & - &  - & - & - & -\\
MMT/MMRS$_H$ & $20.40\pm0.08$ & - & - & - & - & - & -\\
MMT/MMRS$_K$ & $20.51\pm0.08$ & - & - & - & - & - & -\\
WISE$_{\rm W1}$ & - & - & $16.65\pm0.07$  &  - & - & - & -\\
WISE$_{\rm W2}$ & - & - & $16.08\pm0.14$  &  - & - & - & -\\
WISE$_{\rm W3}$ & - & - & $12.42\pm0.49$  &  - & - & - & -\\
WISE$_{\rm W4}$ & - & - & $9.01$  &  - & - & - & -\\
\hline 
\enddata 
\end{deluxetable*} 
\begin{figure*}
  \centering 
  \subfloat {\includegraphics[scale=0.50]{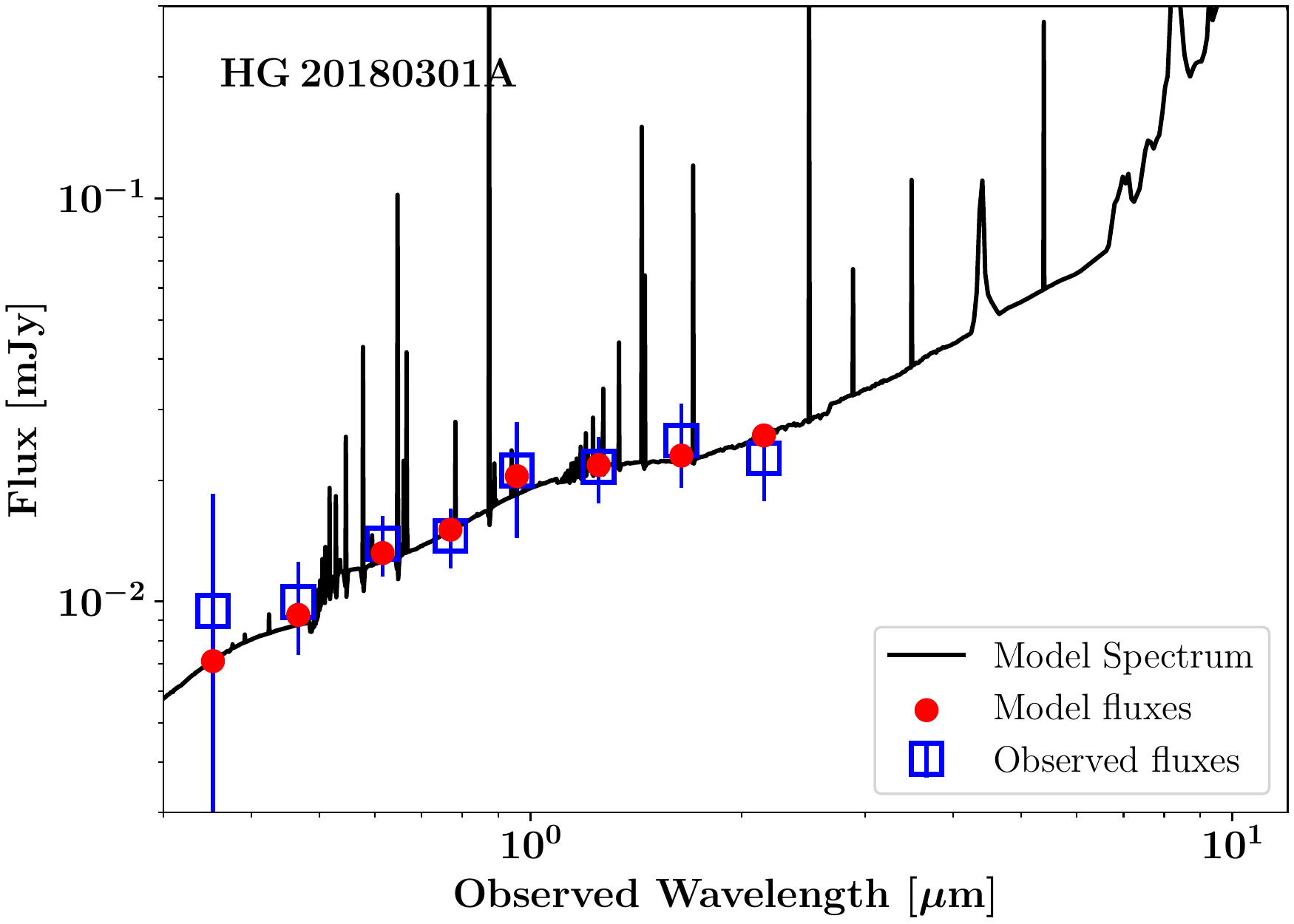}}
  \subfloat {\includegraphics[scale=0.50]{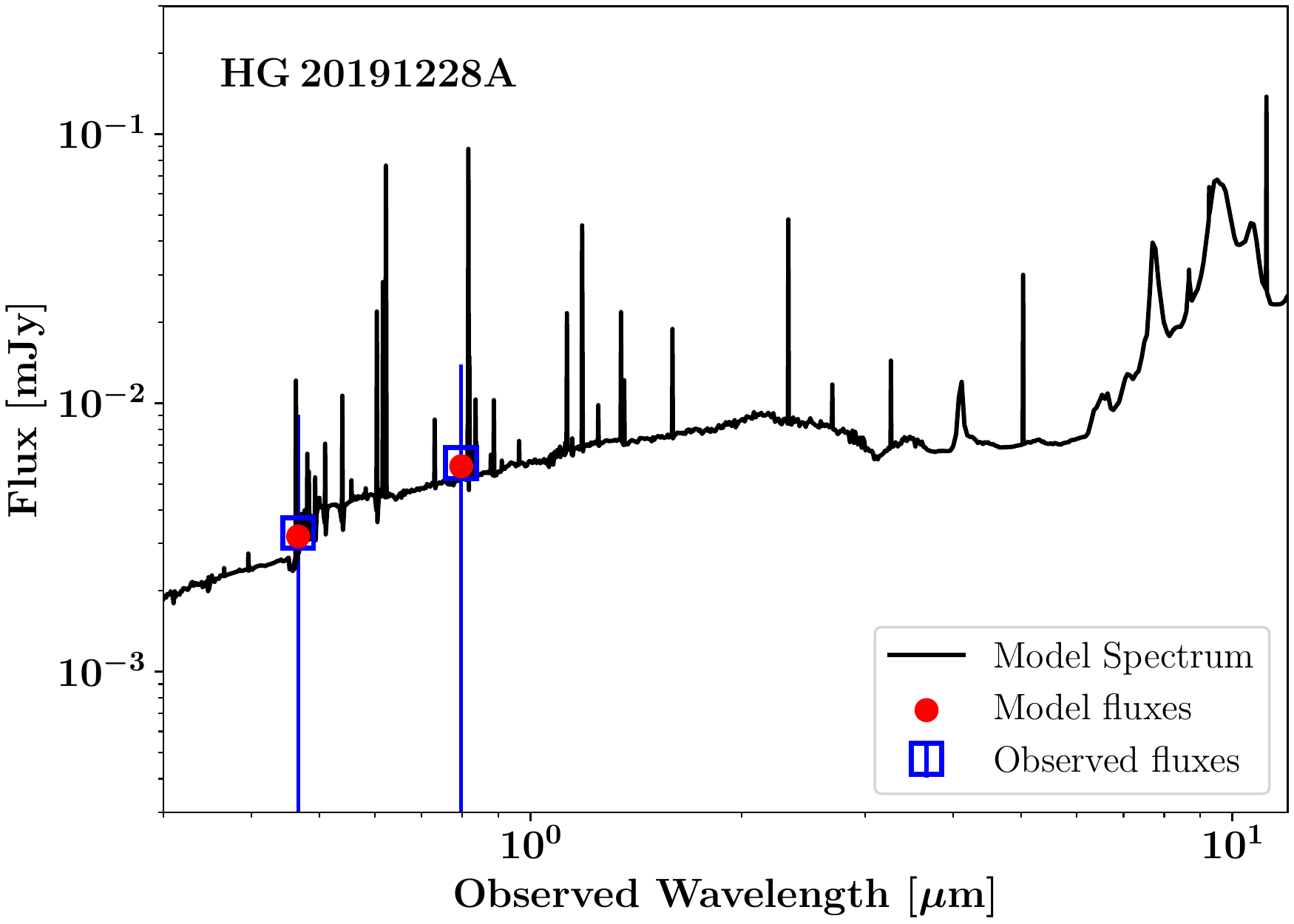}}
  \qquad 
  \subfloat {\includegraphics[scale=0.50]{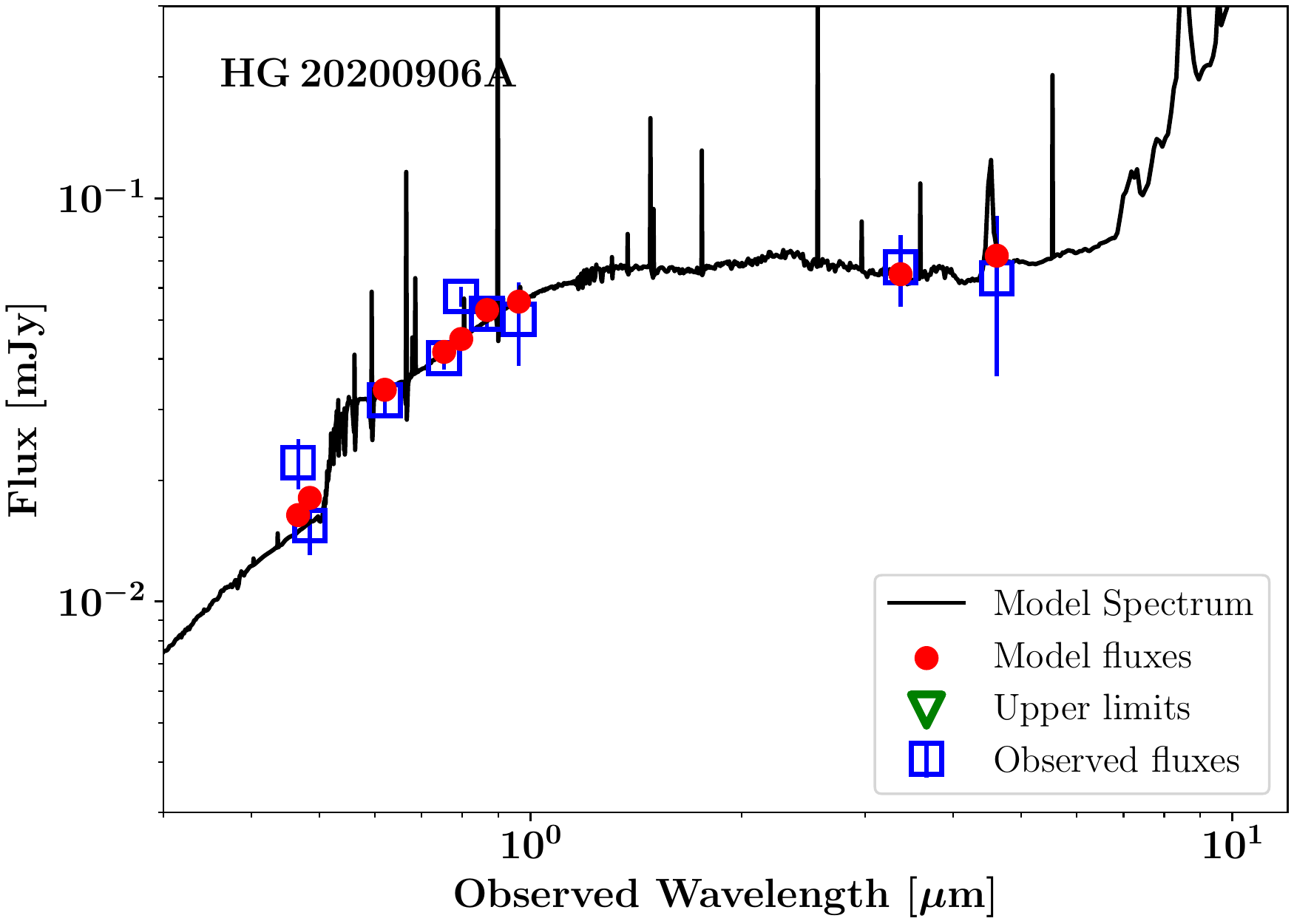}}
  \subfloat{\includegraphics[scale=0.50]{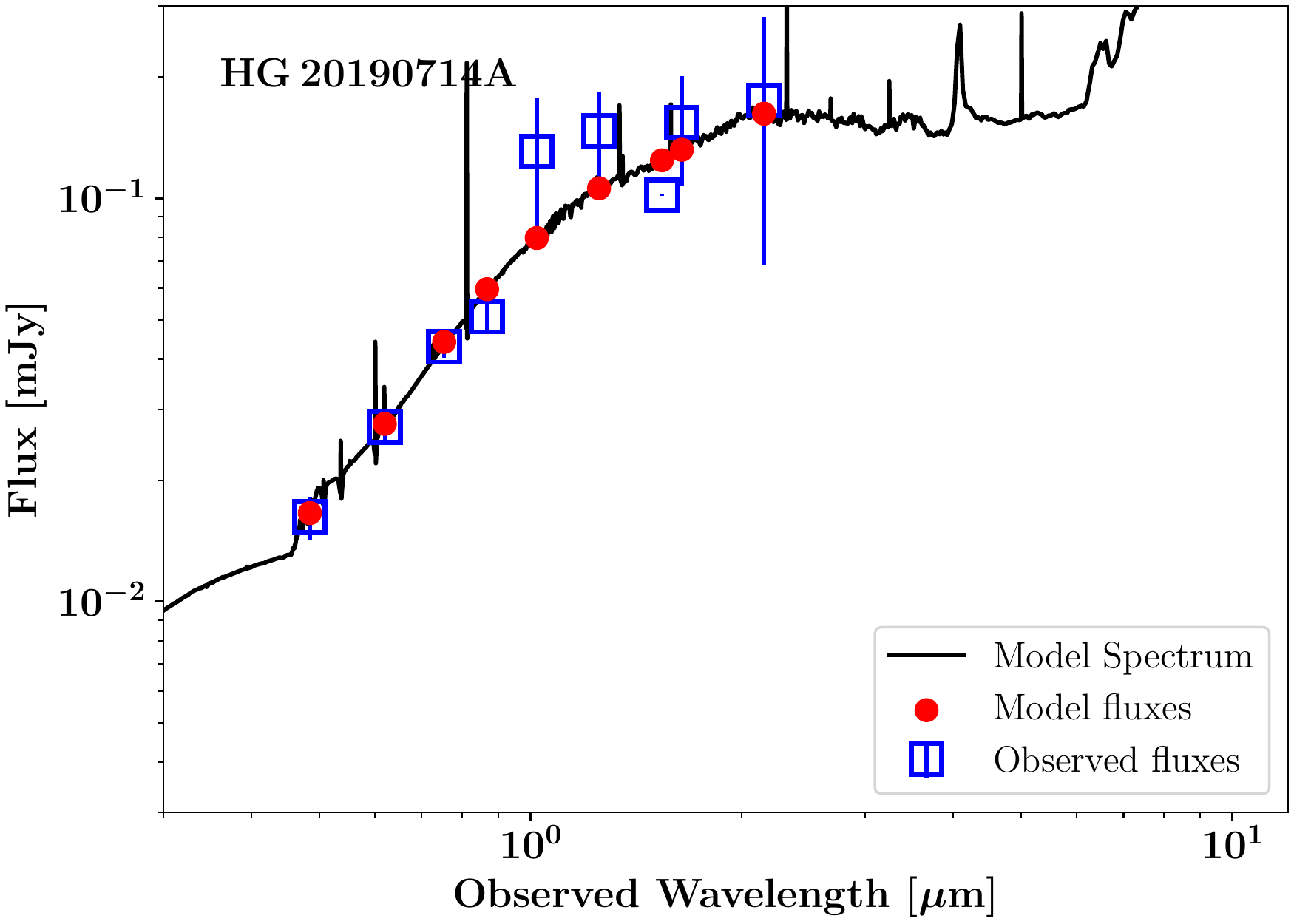}}
  \caption{SED models for the host galaxies of FRB20180301A, FRB20191228A, FRB20200906A, FRB20190714A. The best-fit SED models from CIGALE are shown as solid black lines, the observed magnitudes (corrected for Galactic extinction and converted into fluxes) as blue squares, and the model fluxes as red dots. In all models, the redshift has been fixed to the redshift of respective host galaxies.}
  \label{fig:cigale}
\end{figure*}

\begin{figure*}
  \ContinuedFloat 
  \centering 
  \subfloat{\includegraphics[scale=0.50]{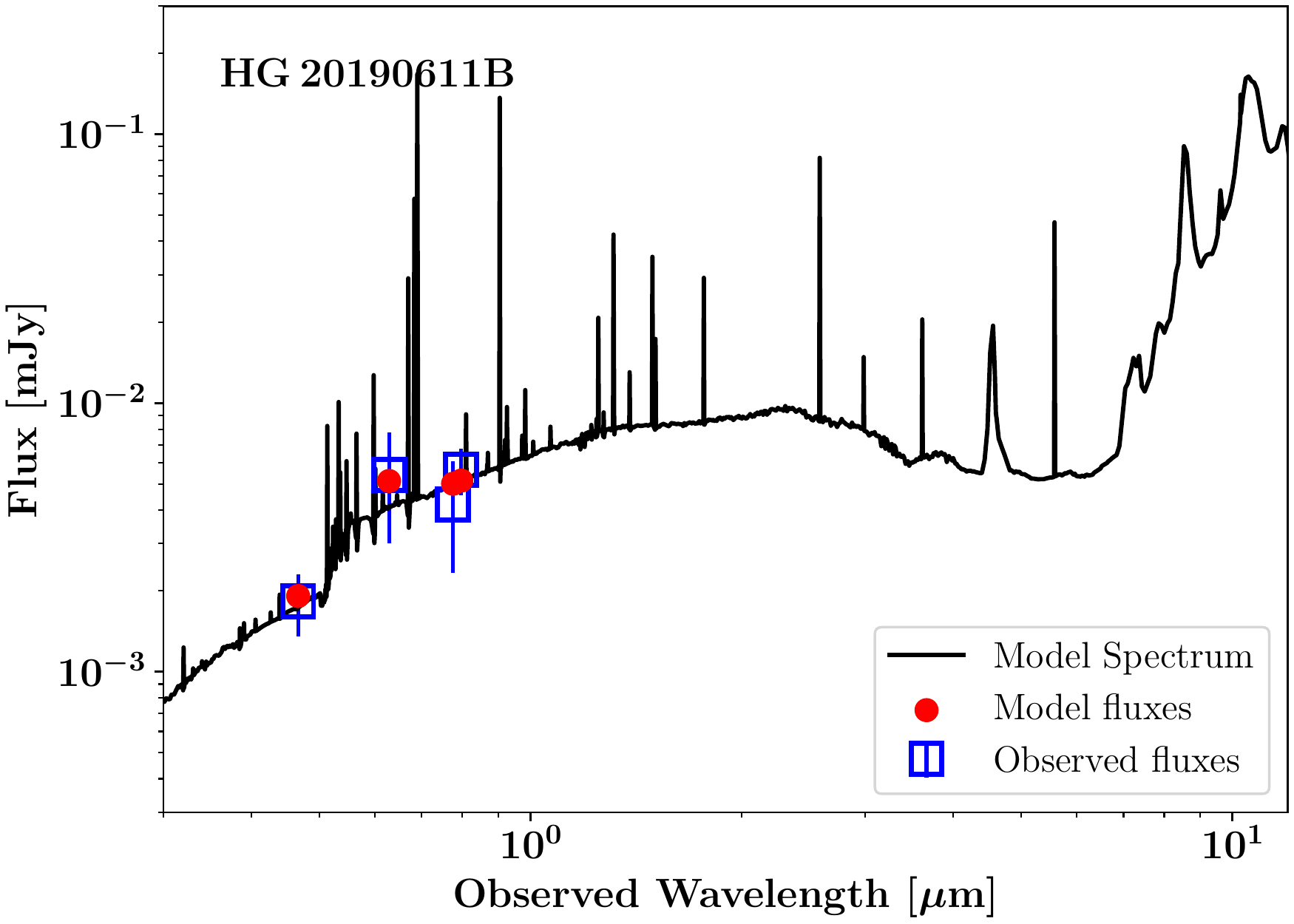}}
  \subfloat{\includegraphics[scale=0.50]{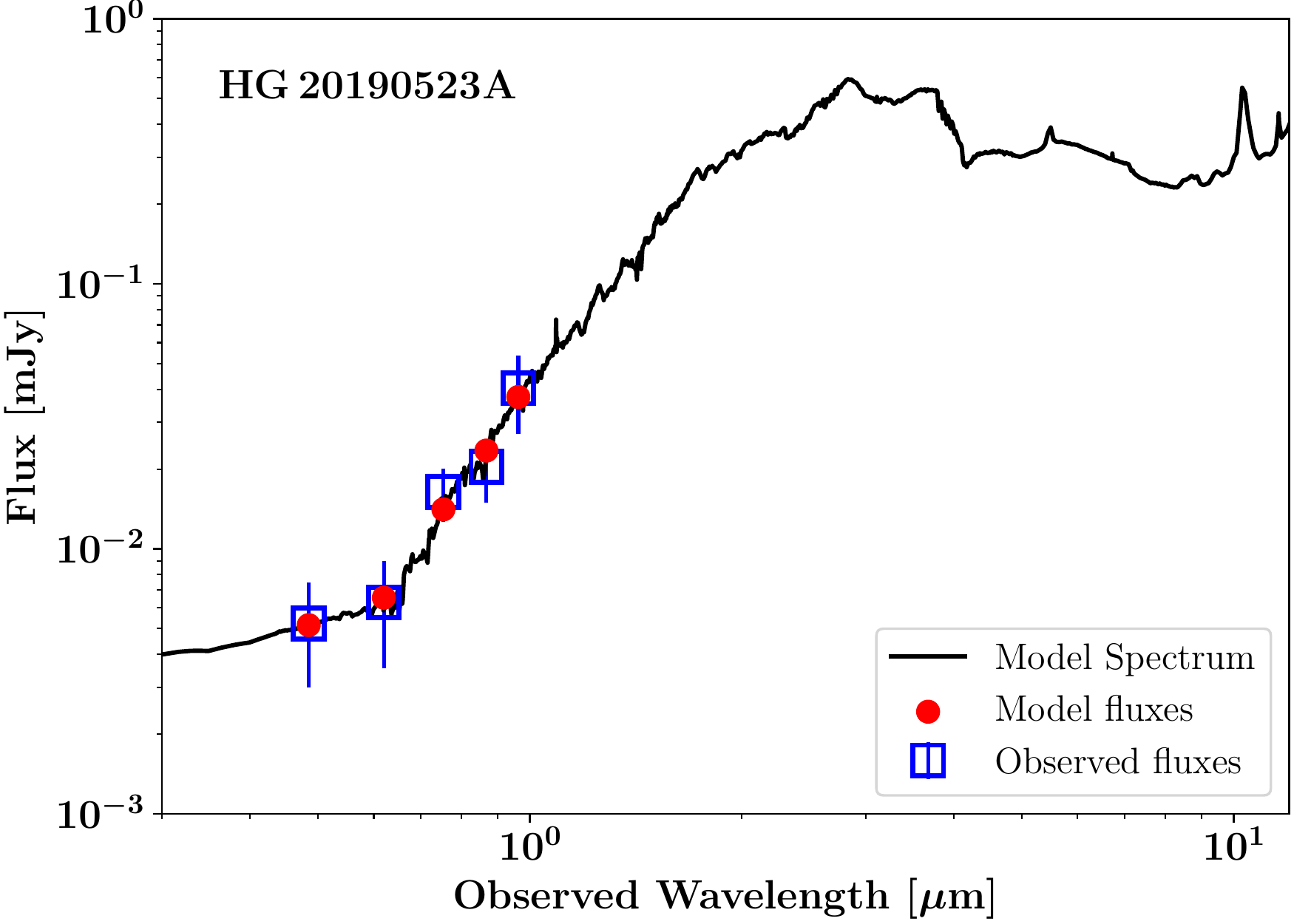}}
  \qquad
  \subfloat{\includegraphics[scale=0.50]{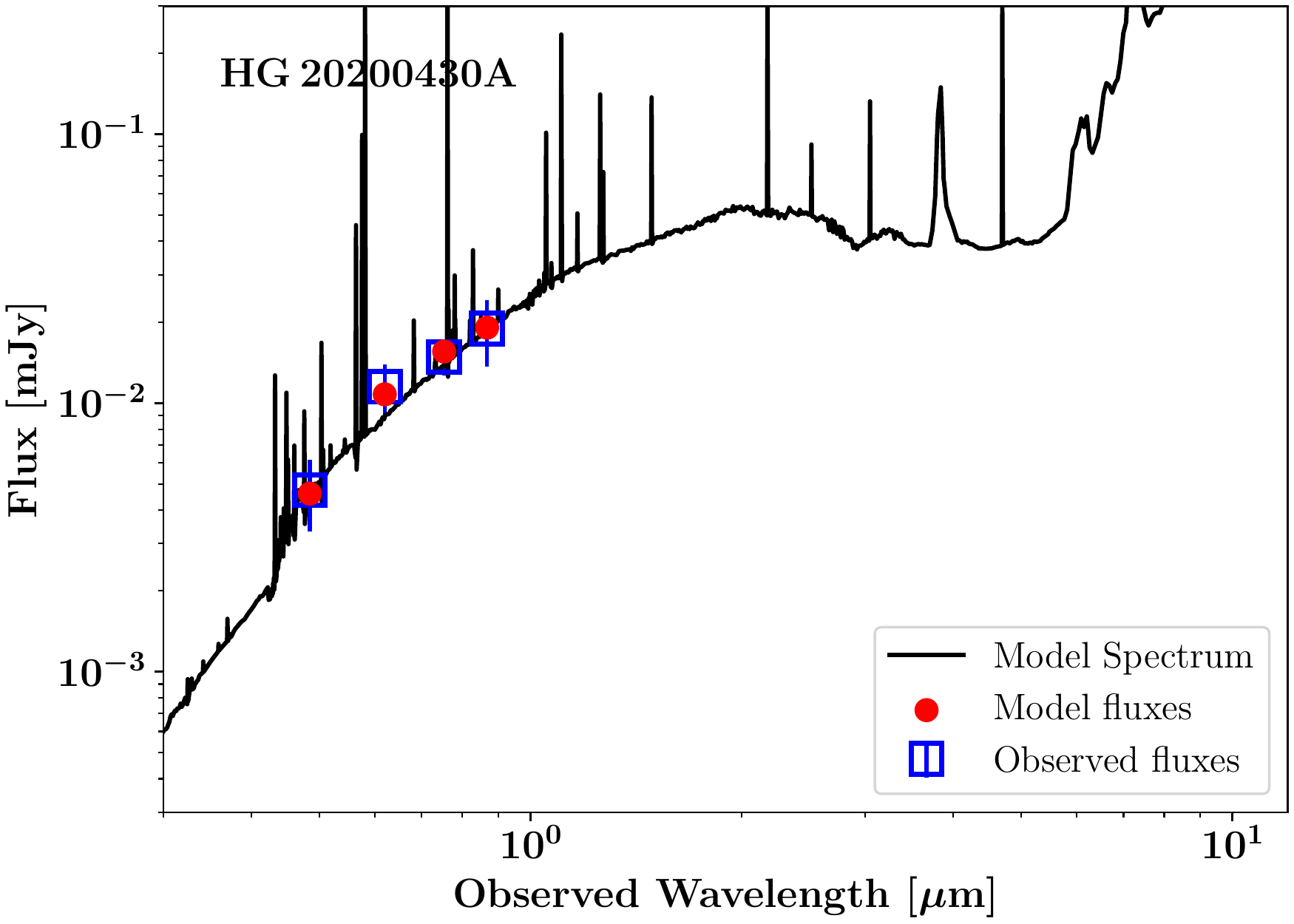}}
  \caption{SED models for the host galaxies of FRB20190611B, FRB20190523A and FRB20200430A.}
  \label{fig:cont}
\end{figure*}

\end{document}